\documentclass[twocolumn,iop,aps,superscriptaddress,showpacs,floatfix]{revtex4-1}

\usepackage[colorlinks=true,citecolor=myblue,linkcolor=myblue,urlcolor=myblue]{hyperref}
\usepackage{mathptmx}
\usepackage{subfigure}
\usepackage{dcolumn}
\usepackage{amsmath,amssymb}
\usepackage{bm}
\usepackage{color}
\usepackage{latexsym}
\usepackage{epstopdf}
\usepackage{color}
\usepackage[english]{babel}
\usepackage{latexsym}
\usepackage{graphicx}
\usepackage{appendix}

\usepackage{psfrag,graphicx} 
\usepackage{epsf} 
\usepackage{subfigure} 
\usepackage{amsfonts}
\usepackage{bm}
\usepackage{natbib}
\usepackage{epstopdf}
\usepackage{appendix}
\usepackage{changes}
\usepackage{bbm}
\usepackage[export]{adjustbox}

\definecolor{mygrey}{gray}{0.35}
\definecolor{myblue}{rgb}{0.2,0.2,0.8}
\definecolor{myzard}{cmyk}{0,0,0.05,0}
\definecolor{mywhite}{rgb}{1,1,1}
\definecolor{myred}{rgb}{1,0.,0.3}

\def\be{\begin{equation}}
\def\ee{\end{equation}}
\def\ba{\begin{align}}
\def\enda{\end{align}}
\def\bi{\begin{itemize}}
\def\ei{\end{itemize}}

 \def\ee{\mathord{\rm e}}

 \def\ee{\mathord{\rm e}}

\renewcommand{\ee}{{\rm e}}

\def\beq{\begin{equation}}
\def\beq{\begin{equation}}
\def\eeq{\end{equation}}

 \newcommand{\ket}[1]{|#1\rangle}
 
 \newcommand{\braket}[2]{\langle #1|#2\rangle}

\begin{document}

\title[Short Title]{Quantum control and sensing of nuclear spins by electron spins under power limitations}
\author{Nati Aharon}
\affiliation{Racah Institute of Physics, The Hebrew University of Jerusalem, Jerusalem 
91904, Givat Ram, Israel}
\author{Ilai Schwartz}
\affiliation{NVision imaging, Ulm D-89069, Germany}
\author{Alex Retzker}
\affiliation{Racah Institute of Physics, The Hebrew University of Jerusalem, Jerusalem 
91904, Givat Ram, Israel}
\date{\today}


\begin{abstract}
State of the art quantum sensing experiments targeting frequency measurements or frequency addressing of nuclear spins require to drive the probe system at the targeted frequency. In addition, there is a substantial advantage to perform these experiments in the regime of high magnetic fields, in which the Larmor frequency of the measured spins is large.
In this scenario we are confronted with a natural challenge of controlling a target system with a very high frequency when  the probe system cannot be set to resonance with the target frequency. In this contribution we present a set of protocols that are capable of confronting this challenge, even at large frequency mismatches between the probe system and  the target system, both for polarisation and for quantum sensing.

\end{abstract}
\maketitle

\begin{figure}[t]
\begin{center}
\includegraphics[width=0.48\textwidth]{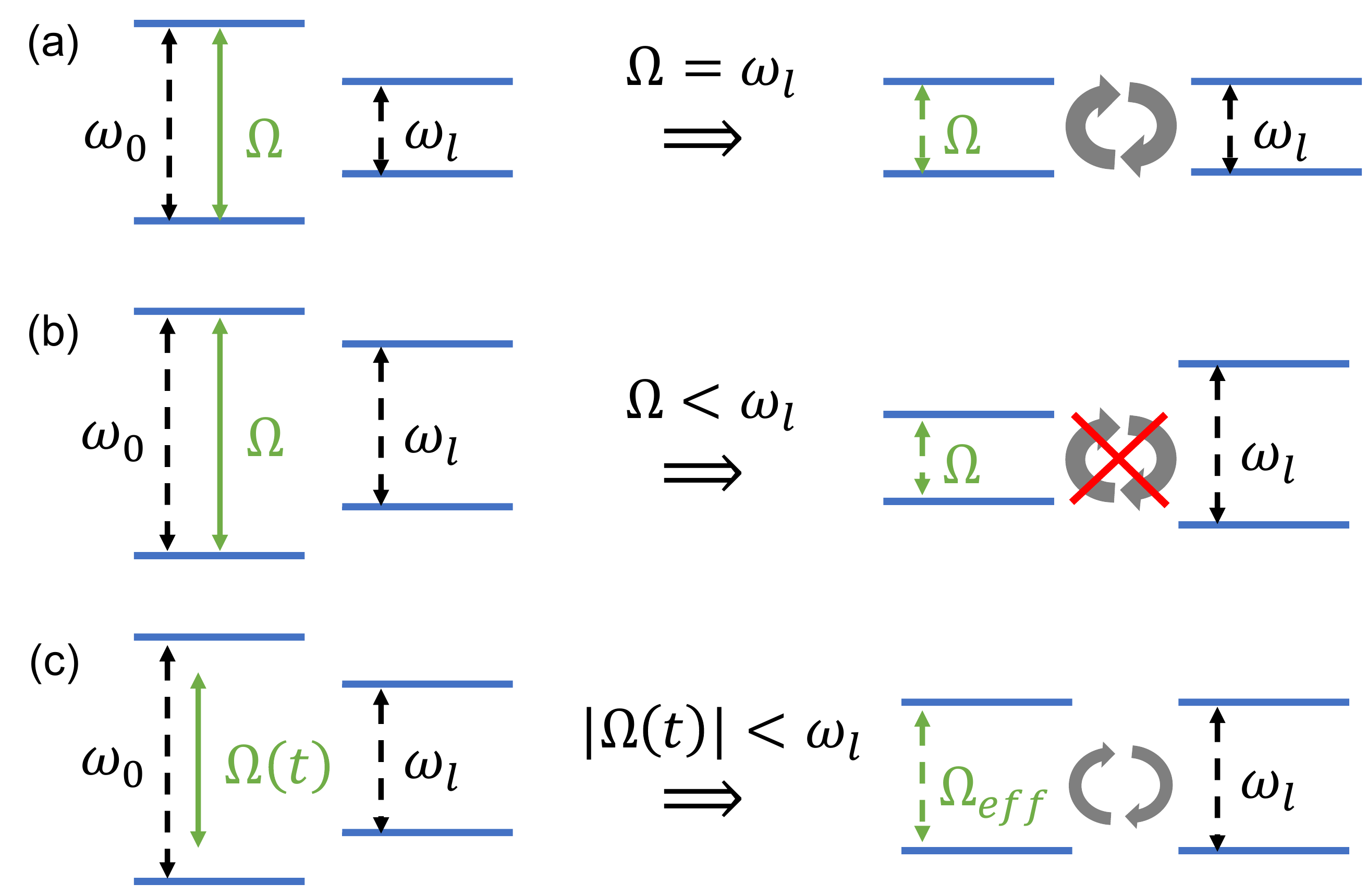}
\end{center}
\caption{The main problem. (a) Control and sensing of nuclear spins is achieved by satisfying the HH condition. The  electron is driven with a RF ($\Omega$) that is equal to the nuclear LF ($\omega_l$). This results in dressed electron states that are on resonance with the LF, enabling the electron-nucleus spin interaction. (b) The electron is driven with a bounded RF, which is smaller than the LF ($\Omega<\omega_l$) and thus no coupling can be achieved. This is a typical problem in the high magnetic fields regime. (c) We propose a set of protocols where even though the electron spin is driven with a bounded RF, $|\Omega(t)|<\omega_l$, an effective dressed electronic energy gap that is equal to the  LF is obtained.  The effective electron-nucleus coupling strength decreases for a larger frequency mismatch $\omega_l -\Omega$. Doted lines (solid lines) indicate energy gaps (driving fields).}
\label{idea3}
\end{figure}

\emph{Introduction ---}
Nuclear spins control by electrons is a ubiquitous in quantum technology setups. Control experiments of nuclei in solids were realized via defects in diamond \cite{jelezko2006single,lee2013readout}, especially  NV centers in diamond \cite{neumann2010single,pfender2018back,unden2016quantum,jiang2009repetitive,dutt2007quantum},  Silicon Carbide \cite{falk2015optical,ivady2015theoretical} and Silicon \cite{morton2008solid,pla2013high}.
These experiments were motivated by quantum computing \cite{yao2012scalable,childress2013diamond,robledo2011high,van2012decoherence,taminiau2014universal}, quantum sensing  \cite{aslam2018nanoscale,perunicic2014towards,kong2017atomic} and dynamical nuclear polarization \cite{shagieva2018microwave,pagliero2018multispin,broadway2018quantum,fernandez2018toward,scheuer2016optically,chen2016resonance}. 
Nuclear spins control requires to work at resonance, which is manifested by the Hartmann-Hahn (HH) condition \cite{hartmann1962nuclear}.
The HH condition requires to equate the Rabi frequency (RF) at which the electron is driven to the Larmor frequency (LF) of the nuclei (Fig. \ref{idea3} (a)).  
There is, however, a strong motivation to perform experiments at high magnetic fields due to the prolonged nuclear coherence time and the improvement in single-shot readout.
Such experiments are very challenging and only a few were realized successfully \cite{aslam2015single,haberle2017nuclear,stepanov2015high,pfender2017protecting}.
Moreover, in some experiments (e.g., in biological environments) the maximal RF is restricted by deleterious heating effects that are associated with high power. In such cases it is challenging to reach the high RF that matches the  nuclear LF  (Fig. \ref{idea3} (b)).

In this Letter we present a few schemes that can overcome this limitation in the various regimes of the mismatch between the RF and the targeted LF. We show that by employing a detuned driving field with a constant bounded RF or a driving field with a (bounded) modulated RF or a modulated phase, it is possible to reach the HH condition (Fig. \ref{idea3} (c)). Although such protocols were achieved with pulsed schemes that require high power \cite{casanova2018shaped}, we introduce simpler continuous drive based constructions that are significantly more power-efficient \cite{gordon2008optimal,cao2017protecting}. While we focus on the NV center, the presented schemes are general and applicable to both the optical and microwave domains, and hence to a variety of atomic and solid state systems.

\emph{The model ---}
We consider an NV center electronic spin that is interacting with a single or several nuclei via the dipole - dipole interaction.
Under an on-resonance drive,  the Hamiltonian of the NV and a nuclear spin is given by  \cite{supp}
$H = \frac{\omega_0}{2} \sigma_z + \frac{\omega_l}{2} I_z + g \sigma_z I_x + \Omega_1 \sigma_x \cos \left(\omega_0 t \right),$
where $\omega_0$ corresponds to the NV's energy gap, $\omega_l$ is the nucleus LF,  $\sigma_z$ and $I_z$ are the Pauli operators in the direction of the static magnetic field of the NV and the nucleus respectively, $g$ is the NV - nucleus coupling strength, and $\Omega_1$ is the RF  of the NV drive.
For sensing and control of the nucleus by the NV the HH condition, $\Omega_1 = \omega_l$, must be fulfilled (Fig. \ref{idea3} (a)) \cite{supp}.


In the high magnetic field regime the nuclear LF, $\omega_l = \gamma_n B$, where $\gamma_n$ is the nuclear gyromagnetic ratio and $B$ is the static magnetic field, can be as high as $\sim100$ MHz. Hence, because of either technical limitations or avoidance of heating effects that occur due to the high power that is required to generate such a large RF, it is impossible to fulfil the HH condition by an on-resonance drive. Namely, we must work in the regime where $|\Omega_1|<\omega_l$ (Fig. \ref{idea3} (b)). We term the frequency difference, $\omega_l-\Omega_1$, as the frequency mismatch between the NV  frequency ($\Omega_1$) and the nuclear LF ($\omega_l$). 

We propose a set of protocols where even though the electron is driven with a bounded RF, $|\Omega(t)|<\omega_l$, an effective dressed electronic energy gap that is equal to the  LF is obtained, and hence, the resonance condition is retrieved. 
Most generally, we consider the Hamiltonian 
$ H_s = \frac{\omega_0}{2} \sigma_z + \frac{\omega_l}{2} I_z + g \sigma_z I_x + \Omega_1(t) \sigma_x \cos \left(\phi(t)  \right), $
where $ \Omega_1(t) $ and  $\phi(t)$ are the modulated RF and modulated phase of a general driving field. 
The functions $\phi(t)$ and $\Omega_1(t)$ are our  control tools that are used in order to reach the resonance condition in the small and large frequency mismatch regimes respectively,  and therefore enable to probe the nuclei parameters and polarize it in the high magnetic field regime.

\emph{Small frequency mismatch ---}
In continuous dynamical decoupling it is more beneficial to rely on a control by a robust phase modulation (PM) than on a control by a noisy amplitude modulation (AM) \cite{cohen2017continuous}. This concept was verified experimentally \cite{farfurnik2017experimental,cao2017protecting} and here we further develop it to design efficient and robust control in the high magnetic field regime when the frequency mismatch is small. 
This scenario is relevant for a LF of $\sim 1-10$ MHz. For example, the  LF of $^{13}$C  ($^{15}$N) at a magnetic field of $1$T ($1.5$T) is $10$ MHz ($6.5$ MHz).
There are two key advantages of PM. First, PM is much more stable than a noisy AM and therefore results in longer coherence times. Second, the extra frequency that is required to fulfil the resonance condition ($\omega_l-\Omega_1$) originates only from the PM and therefore does not require extra power beyond the power limit of the bounded RF $\Omega_1$ \cite{supp}.


We consider the following Hamiltonian of the NV and the nucleus, 
$H = \frac{\omega_0}{2} \sigma_z + \delta B(t) \sigma_z+ \frac{\omega_l}{2} I_z + g \sigma_z I_x 
 + \left( \Omega_1 + \delta \Omega_1(t)  \right) \sigma_x \cos \left( \omega_0 t  + 2 \frac{\Omega_2}{\Omega_1} \sin(\Omega_1 t)  \right), $
where $ \delta B(t) $ is the magnetic noise,  $\Omega_1$  is the RF of the drive, which defines the PM according to $\phi\left(t\right)=2 \frac{\Omega_2}{\Omega_1} \sin\left(\Omega_1 t\right) $, and $ \delta \Omega_1(t) $ is the amplitude noise in  $\Omega_1$. The NV dynamics is modulated by two frequencies, $\Omega_1$ and $\Omega_2$, and thus we may expect transitions to occur whenever the resonance condition, $\Omega_1 + \Omega_2 = \omega_l$ is met.
Indeed, this Hamiltonian results in double-dressed NV states for which we have that \cite{supp}
$H_{II} \approx \frac {\Omega_2}{2}\sigma_z+ \frac{\omega_l}{2} I_z 
- \frac{g}{2} \left(  \sigma_{+} \left(e^{i \Omega_1 t}- e^{-i \Omega_1 t}\right)+ \sigma_{-} \left(e^{-i \Omega_1 t} - e^{i \Omega_1 t}\right) \right) I_x, $
where $H_{II}$ is the Hamiltonian in the second interaction picture (IP) and in the basis of the double-dressed states. From this expression it is seen that a resonance condition appears when $\Omega_1 +\Omega_2  = \omega_l$ (or when  $\Omega_1 -\Omega_2  = \omega_l$).
Even though the power of the driving field is $\propto\Omega_1^2$ and is independent of $\Omega_2,$ higher Larmor frequencies than what is available by the peak power in a common HH scheme are reachable. 
While the modulation by the frequency $\Omega_1$ originates from  AM and requires a power  of $\propto\Omega_1^2$, the second modulation by the frequency $\Omega_2$ originates from the PM and as such it is not associated with  extra power. Specifically, for $\Omega_2=\Omega_1$ the ratios of the peak power (the maximal instantaneous power value)  and the cycle power (the power that is required for a complete energy transfer (flip-flop) between the NV and the nucleus) between a common HH drive and a phase modulated drive are $4$ and $2$ respectively  \cite{supp}.  
Moreover, PM may result in significantly prolonged coherence times due to the precise phase control of microwave sources, and the elimination (to first order) of amplitude fluctuations in $\Omega_1$ \cite{supp}. 

The above procedure is correct in the limit of $\Omega_2 \ll \Omega_1$. However, we aim to increase $\Omega_2$ as much as possible without reducing the sensitivity. 
To this end, we have to take into account the Bloch-Siegert Shift (BSS) due to the counter-rotating terms of the second modulation $\Omega_2$, which induces a shift of the resonance. In addition,  this decreases the coupling to the nucleus, and more importantly, the coherence time of the NV as the decoupling effect of the drive is not effective any more (Fig. \ref{Coherence} (blue)). 
To improve this, we suggest to correct the BSS when adjusting the frequency $\Omega_1$ in the PM $\phi\left(t\right)=2 \frac{\Omega_2}{\Omega_1} \sin\left(\Omega_1 t\right) $ and modify it to 
$\tilde{\Omega}_1 = \frac{1}{3}\left( \Omega_1+ \sqrt{4 \Omega_1^2+3 \Omega_2^2}\right)$. In this case, the resonance frequency is $\tilde{\Omega}_1 +\tilde{\Omega}_2  = \omega_l$, where $\tilde{\Omega}_2 = \frac{\Omega_2}{2}\left(1+\frac{\Omega_1+\tilde{\Omega}_1}{\sqrt{\Omega_2^2+\left(\Omega_1+\tilde{\Omega}_1\right)^2}}\right)$  \cite{supp}.

\begin{figure}[t]
\begin{center}
\includegraphics[width=0.48\textwidth]{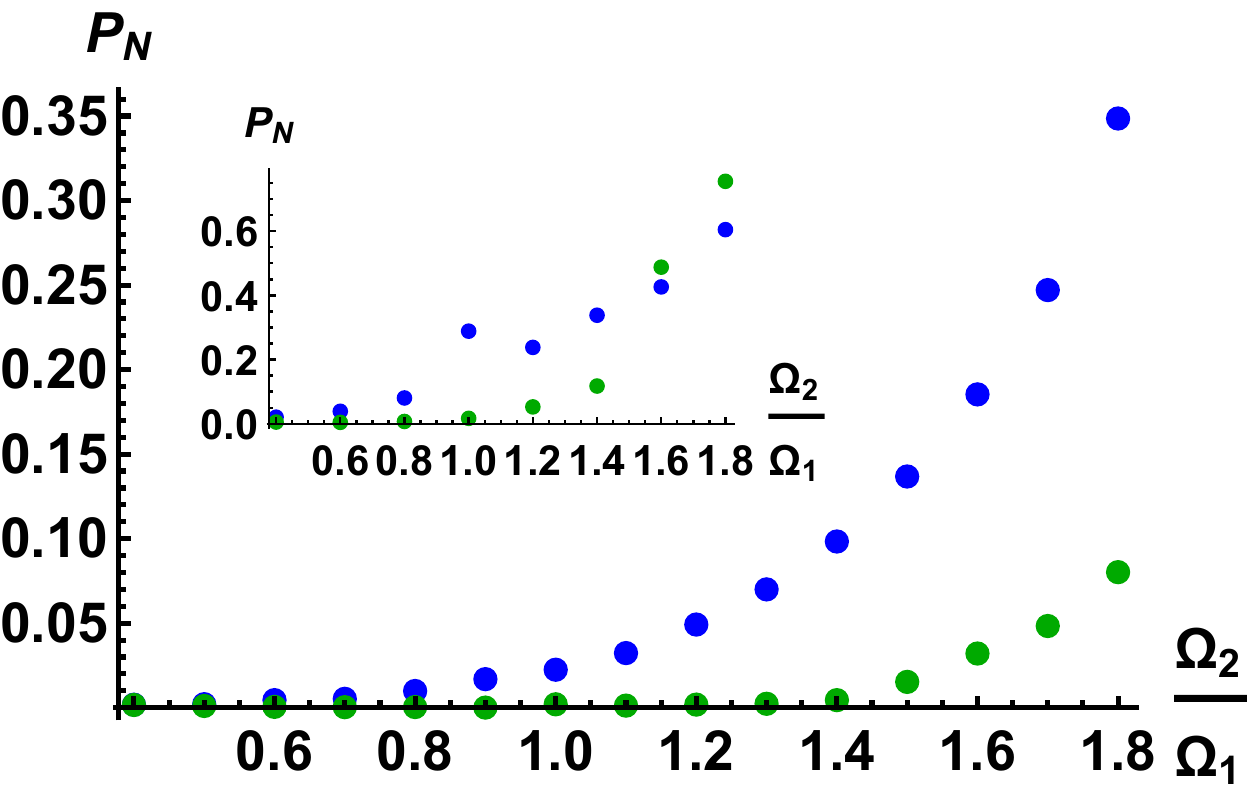}
\end{center}
\caption{
Polarization as a function of $\frac{\Omega_2}{\Omega_1}$ in the strong and weak (inset) coupling regimes  without BSS correction (blue) and with BSS correction (green). Strong coupling regime: without correction the polarization rate begins to sharply decrease at $\Omega_2 \approx \Omega_1$. The  correction enables to maintain good polarization rates up to $\Omega_2 \approx 1.8\Omega_1$. Weak coupling regime: The analysis takes noise into account. The polarization is effective up to $\Omega_2 \approx 1.4 \Omega_1$.
}
\label{Polar}
\end{figure}

\begin{figure}[t]
\begin{center}
\includegraphics[width=0.48\textwidth]{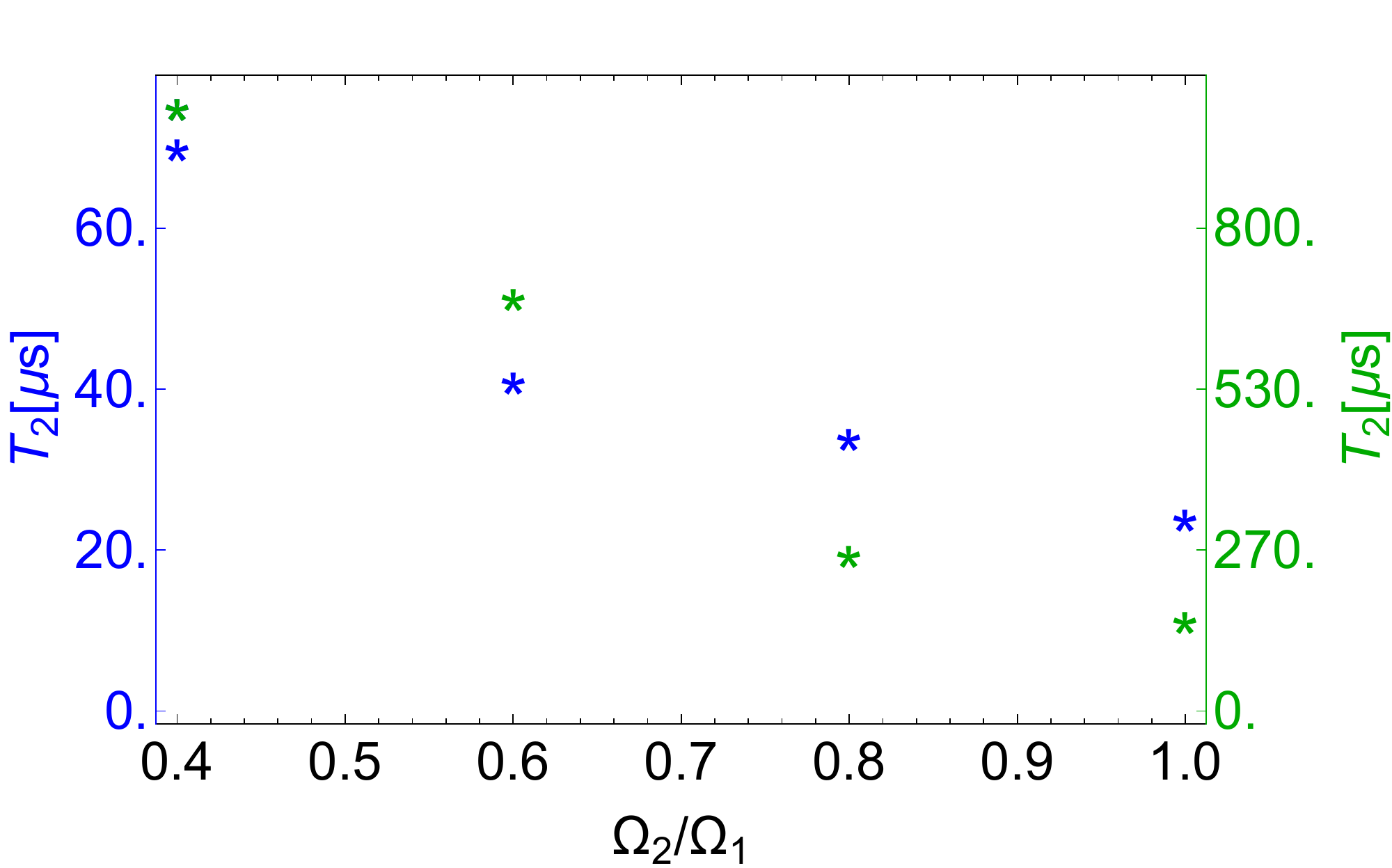}
\end{center}
\caption{Coherence time ($T_2$) as a function of $\frac{\Omega_2}{\Omega_1}$. Without the BSS correction  (blue) - at the regime of an efficient polarization, $T_2$ is decreased as $\Omega_2$ is increased. The optimal $T_2$ is sharply peaked at $\frac{\Omega_2}{\Omega_1}\approx0.125$ with $T_2\approx 330\:\mu s$ (not shown). With the BSS correction (green) - a long $T_2$ time is maintained while increasing $\Omega_2$. The optimal $T_2$ is  peaked at $\frac{\Omega_2}{\Omega_1}\approx0.4$ with $T_2\approx 1000\:\mu s$. The coherence time is a crucial parameter in the efficiency of control and estimation.}
\label{Coherence}
\end{figure}

In Fig. (\ref{Polar}) we show simulation results \cite{supp} for the nucleus polarization as function of $\frac{\Omega_2}{\Omega_1}$. In the main figure we consider the strong coupling regime, where  the polarization time $t=2\pi/g$ is much shorter than the decoherence time of the NV center and hence, decoherence effects are neglected. In the inset we consider the weak coupling regime where noise decreases the polarization rate  \cite{supp}. We define the nuclear spin polarization, $P_N$, as the probability of the nuclear spin to be in its initial state $\ket{\uparrow_z}$. Specifically, we initialize the NV-Nucleus state to    $\ket{\psi_i}=\ket{\downarrow_z}_{NV}\ket{\uparrow_z}_{N}=\ket{\downarrow_z \uparrow_z}$ and calculate the polarization according to $P_N =|\braket{\uparrow_z \uparrow_z}{\psi}|^2 + |\braket{\downarrow_z \uparrow_z}{\psi}|^2$, where $\ket{\psi}$ is the joint NV-Nucleus state at the optimal polarization time. Hence, $P_N=0$ corresponds to optimal polarization and $P_N=1$ corresponds to no polarization at all. 
While in the strong coupling regime the correction always results in better polarization rates, in the weak coupling regime the advantage of correction is lost at $\Omega_2 \approx 1.5 \Omega_1$. In  Fig. (\ref{Coherence}) we show the expected coherence times, $T_2$, of the NV as function of $\frac{\Omega_2}{\Omega_1}$ \cite{supp}. Without the correction 
the optimal coherence time is sharply peaked at $\frac{\Omega_2}{\Omega_1}\approx0.125$ with $T_2\approx 330\:\mu s$ (not shown).
The coherence time is reduced when $\Omega_2$ is increased due to an amplitude mixing of $\propto \frac{\Omega_2}{\Omega_1}$ between the dressed states, which introduces back a first order contribution of the drive noise $\propto \frac{\Omega_2} {\Omega_1} \delta \Omega_1$.  This decoherence is greatly mitigated by the correction of the BSS up to  $\Omega_2 \approx  \Omega_1$, which results in an improvement of one order of magnitude in the coherence times. With the correction 
the optimal coherence time is  peaked at $\frac{\Omega_2}{\Omega_1}\approx0.4$ with $T_2\approx 1000\:\mu s$. In this case, the coherence time is mainly limited by the second order contribution of the drive noise  $\sim\frac{\delta \Omega_1^2}{\Omega_2}$. The BSS correction enables to further increase $\Omega_2$ and  results in  prolonged NV's coherence times and higher polarization rates.

\emph{Large frequency mismatch ---}
The natural way to compensate for the frequency mismatch  is to introduce a detuning ($\delta$) to the drive. This detuning induces an extra modulation that creates an effective frequency of  $\sqrt{\Omega_1^2 + \delta^2}$, which in principle, can be as high as needed $\left(\sqrt{\Omega_1^2 + \delta^2} \gg \Omega_1\right)$. When the effective frequency $\sqrt{\Omega_1^2 + \delta^2}$ is equal to the  LF, the HH condition is fulfilled and the electron-nucleus interaction is enabled \cite{supp}. This however, comes with a price; the electron-nucleus coupling strength is decreased by a factor of $\sim\frac{\Omega_1}{\delta}$ \cite{supp} (Fig. \ref{idea3} (c)). 
Here the decoupling effect of a resonant drive vanishes and the NV's coherence time  approaches  $T_2^*$. In \cite{supp} we show how to circumvent this by adding a second drive. 
This scheme, however, could be extremely power efficient, e.g., for $\delta=10 \Omega_1$ the ratios of the peak power and the cycle power between a common HH drive and a detuned drive are $101$ and $10.1$ respectively  \cite{supp}.

\begin{figure}[t!]
\includegraphics[width=0.48\textwidth]{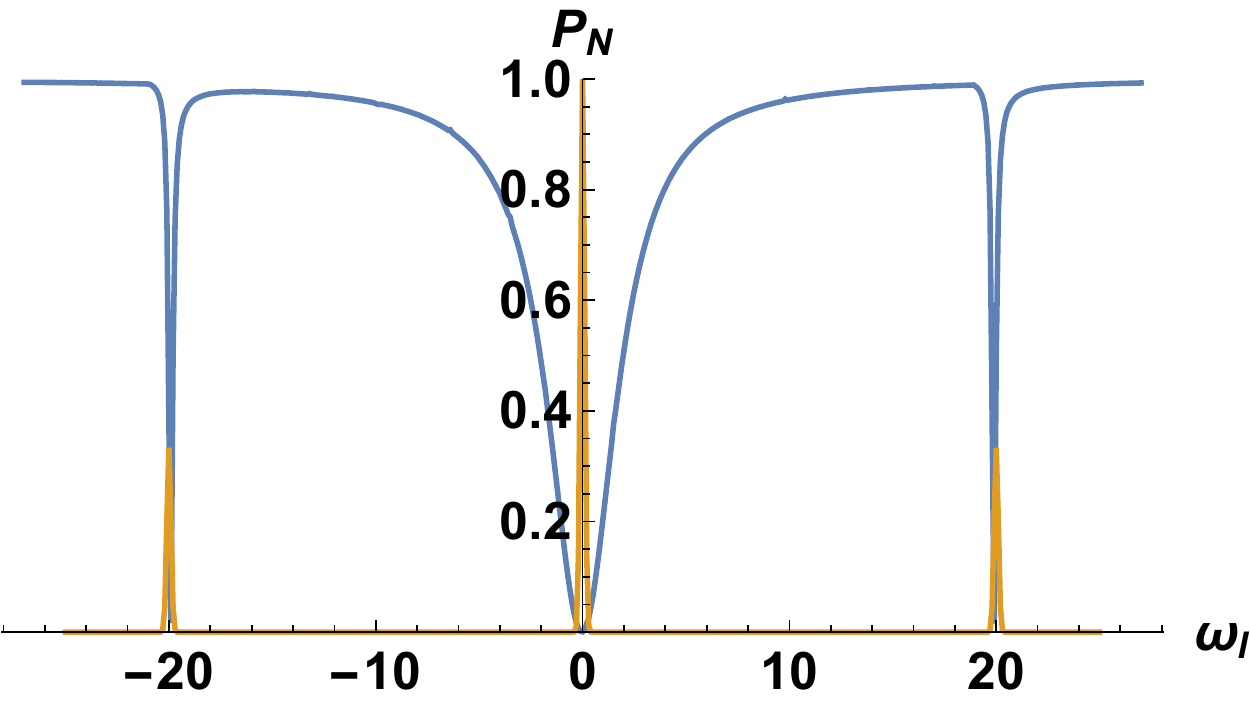}
\caption{Polarization as a function of the nuclear LF in units of tenth of the coupling strength (blue line). The central resonance corresponds to  $\omega_l=\Omega_0$. The two  sidebands correspond to $\omega_l = \Omega_0 \pm \Omega_2.$ The $y$ axis corresponds to the occupation of the nucleus when the initial state is the $\vert \uparrow_z \rangle$ state, i.e., $P_N = 1.$ The main deep is broader than the side deeps because the coupling at the sideband frequencies is reduced.  
In contrast, the yellow line represent the analysis of the quantum sensing Hamiltonian, which is much narrower and is not limited by the coupling strength. The numerical simulations were performed with  $\Omega_0 = 1.5$ MHz, $\Omega_1=0.1$ MHz, $\Omega_2=1$ MHz, and $g= 0.05 \Omega_2$.}
\label{Bessel3}
\end{figure}

An alternative way to reach the resonance is to modulate the amplitude of the drive. This AM generates higher harmonics of the modulation frequency  that  can be tuned to be on-resonance with the LF. 
We start with the Hamiltonian
$ H = \frac{\omega_0}{2} \sigma_z + \frac{\omega_l}{2} I_z + g \sigma_z I_x + \Omega(t) \sigma_x\cos(\omega_0 t) $ 
and set $\Omega(t) = \Omega_0 + \Omega_1 \cos(\Omega_2 t)$. 
Moving to the IP with respect to $H_0 =  \frac{\omega_0}{2} \sigma_z$ and making the rotating-wave-approximation (RWA) $\left(\omega_0 \gg   |\Omega(t)|\right)$ we obtain
$ H_I  = \frac{\Omega(t)}{2} \sigma_x + \frac{\omega_l}{2} I_z +  g \sigma_z I_x,  $
which in the basis of the NV dressed states ($x \rightarrow z$, $z \rightarrow -x$, and $y \rightarrow y$) is given by  
$ H_I  = \frac{\Omega(t)}{2} \sigma_z + \frac{\omega_l}{2} I_z -  g \sigma_x I_x. $
We continue by moving to the second IP with respect to $H_0 = \frac{\Omega(t)}{2} \sigma_z + \frac{\omega_l}{2} I_z$, which results in
$ H_{II} = -g \left( \sigma_+ e^{i \left(  \Omega_0 t + \frac{\Omega_1}{\Omega_2} \sin(\Omega_2 t)  \right) }   +h.c \right)  \left( I_+ e^{i \omega_l t} + I_- e^{-i \omega_l t}  \right). $
The exponent $e^{i \left(  \Omega_0 t + \frac{\Omega_1}{\Omega_2} \sin(\Omega_2 t)  \right)}$ contains the higher harmonics of $\Omega_2,$ i.e., $n \Omega_2$, where $n$ is an integer. This can be seen by the equality 
$ e^{i \left(  \Omega_0 t + \frac{\Omega_1}{\Omega_2} \sin(\Omega_2 t)  \right)} = e^{i \Omega_0 t} \sum_{n = -\infty}^{n = +\infty}\left( i^n J_n\left(\frac{\Omega_1}{\Omega_2} \right) e^{i n \Omega_2 t}  + h.c.\right). $
We can therefore set the resonance condition to  $\Omega_0 + \Omega_2 = \omega_l$. Assuming the RWA $\left(\Omega_2 \gg g\right)$ we get that
$ H_{II} \approx g J_1\left(\frac{\Omega_1}{\Omega_2} \right) \left( i \sigma_+ I_- -i \sigma_- I_+  \right) $
when the resonance condition is fulfilled. 
In the regime of $\Omega_2 \gg \Omega_1$,  $J_1\left( \frac{\Omega_1}{\Omega_2}  \right)\approx \frac{\Omega_1}{2\Omega_2}.$  Hence, the coupling strength is similar to the one in the previous method, however, this scheme is robust to magnetic noise. Numerical analysis of this method is shown in Fig. \ref{Bessel3}.
With  a single AM the method suffers from amplitude fluctuations in $\Omega_0$, 
which could be eliminated by realising this as a second drive from a PM \cite{supp}. This scheme is also power efficient, e.g., for $\Omega_2 = 9 \Omega_0$ the ratios of the peak power and the cycle power between a common HH drive and an amplitude modulated drive are $25$ and $3.7$ respectively  \cite{supp}.

\emph{Quantum sensing ---}
Addressability is the ability of a probe to individually address  and control nuclear spins, which was discussed above. 
However, addressability is not necessary for quantum sensing where, e.g., one is only interested in  estimating the  LF,  as in nano-NMR experiments. The resolution of addressability is defined by the ability to control a nucleus with a given frequency $\omega_l$, while leaving nuclei with different frequencies outside of a frequency width $\Delta \omega$ (centered at $\omega_l$) unaffected.
As shown  in  Fig. \ref{Bessel3} in blue,  the addressability resolution is limited by the coupling strength. This is because that all frequencies within a width of the coupling strength from the resonance will couple to the probe. Hence, the stronger the coupling the worst the resolution is and a larger band of frequencies will be addressed by the probe.

However, when the NV  is used to estimate the LF, one would expect that the stronger the coupling the more information would be acquired; an increased coupling strength should improve the resolution and not limit it. 
The addressability resolution limit could be overcome by designing the Hamiltonian differently.
In cases that control is not necessary, and one is just interested in frequency estimation of the nuclei, methods that are not limited by the coupling strength could be designed. 
The difference between the methods is analogous to the difference between Rabi and Ramsey spectroscopy. While power is a limiting factor in the first method (necessitating weak pulses), it poses no limitation in the second method.

The addressability resolution problem occurs as the NV coupling operator term is a $\sigma_{\pm}$ operator that is in charge of energy transfer. This is crucial for control, however, it is not needed for sensing.
An interaction of the addressability type, $g \left( \sigma_- I_+ + \sigma_+ I_-  \right)$, transfers excitations between the two spins as long as their frequency difference is smaller than the coupling strength $g.$ Thus, the target frequencies within a spread of $g$ are  addressed by the probe. However, an interaction of the type $g \sigma_x \left( I_+ + I_- \right) = g \sigma_x I_x$ could be utilized to estimate the frequencies of the target spins with a resolution that is not limited by the coupling strength \cite{schmitt2017submillihertz,bucher2017high,boss2017quantum}.
This can be achieved  by transforming the $\sigma_-,\sigma_+$ operators into a $\sigma_x$ (or $\sigma_y$) operator, which is doable as $\sigma_\pm = \sigma_x \pm i \sigma_y$ and $\sigma_y$ could be eliminated with a suitable control, for example, by adding a strong $\sigma_x$ drive that will eliminate the $\sigma_y$ part.
For the case of the low frequency mismatch this can be achieved by adding an extra drive on the NV, which rotates at $\Omega_2$ (this amounts to $\Omega_s \cos(\omega_0 t)\cos(\Omega_2 t) \sigma_x$). In \cite{supp} we explicitly show that this results in an Hamiltonian that can be used for sensing the LF, i.e, 
$ H_{I} \approx \frac{g}{4} \sigma_z \left( I_x \cos(\delta t) - I_y \sin(\delta t)   \right), $ where $\delta = \Omega_1 +\Omega_2 -\omega_l.$ 
 As the extra term acts as a spin locking at $\Omega_s,$ the robustness of the methods is preserved.
The classical version of this Hamiltonian was used in \cite{schmitt2017submillihertz,boss2017quantum,bucher2017high,laraoui2013high,zaiser2016enhancing,staudacher2015probing,ajoy2015atomic,rosskopf2016quantum,schmitt2017submillihertz,laraoui2011diamond,pfender2016nonvolatile} where it was shown that the resolution is only limited by the clock and signal coherence times. The resolution obtained by this Hamiltonian, which is the generic sensing Hamiltonian, is only limited by the coherence time of the nuclei and the sensitivity is improved with the coupling  strength \cite{gefen2018quantum}.

The same can be done in the large frequency mismatch regime.
The interaction should be changed from the flip - flop interaction  $g \left( \sigma_+ I_- + \sigma_- I_+  \right)$ to $g \sigma_x I_x$ by adding, for example,  a $\sigma_x$ drive to the modulation. In this case the Hamiltonian is transformed to \cite{supp}
$
H \approx g J_1\left(\frac{\Omega_1}{\Omega_2} \right) \sigma_x \left(I_x \cos(\delta t) - I_y \sin(\delta t)    \right). $
The result of using this Hamiltonian for estimating the nuclei's frequencies is shown in Fig. \ref{Bessel3}.
The yellow line is the Fourier transform of the time series of NV measurements for a scenario in which a few nuclei are present at the three frequencies $\Omega_0,\Omega_0 \pm \Omega_1.$
The width of these peaks (one over the total experiment time) is narrower than the peaks of the control method (blue line), which is limited by the  coupling strength. 

The challenge of controlling and sensing high-frequency nuclei under power limitations of the driving fields was addressed both in the small  and large frequency mismatch regimes.
We have designed schemes that are robust  both to magnetic field fluctuations and RF noise. 
The presented protocols could potentially allow for the realization of experiments in an important regime which is currently out of reach and could considerably simplify state of the art experiments.

We would like to note that during the preparation of this manuscript we became aware of a related independent work by Casanova et al. \cite{casanova2019}.


\emph{Acknowledgements}
A. R. acknowledges the support of ERC grant QRES, project No. 770929, grant agreement No 667192(Hyperdiamond), the MicroQC, the ASTERIQS and the DiaPol project.

\baselineskip=12pt
\bibliography{Bib_Alex}

\begin{thebibliography}{51}%
\makeatletter
\providecommand \@ifxundefined [1]{%
 \@ifx{#1\undefined}
}%
\providecommand \@ifnum [1]{%
 \ifnum #1\expandafter \@firstoftwo
 \else \expandafter \@secondoftwo
 \fi
}%
\providecommand \@ifx [1]{%
 \ifx #1\expandafter \@firstoftwo
 \else \expandafter \@secondoftwo
 \fi
}%
\providecommand \natexlab [1]{#1}%
\providecommand \enquote  [1]{``#1''}%
\providecommand \bibnamefont  [1]{#1}%
\providecommand \bibfnamefont [1]{#1}%
\providecommand \citenamefont [1]{#1}%
\providecommand \href@noop [0]{\@secondoftwo}%
\providecommand \href [0]{\begingroup \@sanitize@url \@href}%
\providecommand \@href[1]{\@@startlink{#1}\@@href}%
\providecommand \@@href[1]{\endgroup#1\@@endlink}%
\providecommand \@sanitize@url [0]{\catcode `\\12\catcode `\$12\catcode
  `\&12\catcode `\#12\catcode `\^12\catcode `\_12\catcode `\%12\relax}%
\providecommand \@@startlink[1]{}%
\providecommand \@@endlink[0]{}%
\providecommand \url  [0]{\begingroup\@sanitize@url \@url }%
\providecommand \@url [1]{\endgroup\@href {#1}{\urlprefix }}%
\providecommand \urlprefix  [0]{URL }%
\providecommand \Eprint [0]{\href }%
\providecommand \doibase [0]{http://dx.doi.org/}%
\providecommand \selectlanguage [0]{\@gobble}%
\providecommand \bibinfo  [0]{\@secondoftwo}%
\providecommand \bibfield  [0]{\@secondoftwo}%
\providecommand \translation [1]{[#1]}%
\providecommand \BibitemOpen [0]{}%
\providecommand \bibitemStop [0]{}%
\providecommand \bibitemNoStop [0]{.\EOS\space}%
\providecommand \EOS [0]{\spacefactor3000\relax}%
\providecommand \BibitemShut  [1]{\csname bibitem#1\endcsname}%
\let\auto@bib@innerbib\@empty
\bibitem [{\citenamefont {Jelezko}\ and\ \citenamefont
  {Wrachtrup}(2006)}]{jelezko2006single}%
  \BibitemOpen
  \bibfield  {author} {\bibinfo {author} {\bibfnamefont {F.}~\bibnamefont
  {Jelezko}}\ and\ \bibinfo {author} {\bibfnamefont {J.}~\bibnamefont
  {Wrachtrup}},\ }\href@noop {} {\bibfield  {journal} {\bibinfo  {journal}
  {physica status solidi (a)}\ }\textbf {\bibinfo {volume} {203}},\ \bibinfo
  {pages} {3207} (\bibinfo {year} {2006})}\BibitemShut {NoStop}%
\bibitem [{\citenamefont {Lee}\ \emph {et~al.}(2013)\citenamefont {Lee},
  \citenamefont {Widmann}, \citenamefont {Rendler}, \citenamefont {Doherty},
  \citenamefont {Babinec}, \citenamefont {Yang}, \citenamefont {Eyer},
  \citenamefont {Siyushev}, \citenamefont {Hausmann}, \citenamefont {Loncar}
  \emph {et~al.}}]{lee2013readout}%
  \BibitemOpen
  \bibfield  {author} {\bibinfo {author} {\bibfnamefont {S.-Y.}\ \bibnamefont
  {Lee}}, \bibinfo {author} {\bibfnamefont {M.}~\bibnamefont {Widmann}},
  \bibinfo {author} {\bibfnamefont {T.}~\bibnamefont {Rendler}}, \bibinfo
  {author} {\bibfnamefont {M.~W.}\ \bibnamefont {Doherty}}, \bibinfo {author}
  {\bibfnamefont {T.~M.}\ \bibnamefont {Babinec}}, \bibinfo {author}
  {\bibfnamefont {S.}~\bibnamefont {Yang}}, \bibinfo {author} {\bibfnamefont
  {M.}~\bibnamefont {Eyer}}, \bibinfo {author} {\bibfnamefont {P.}~\bibnamefont
  {Siyushev}}, \bibinfo {author} {\bibfnamefont {B.~J.}\ \bibnamefont
  {Hausmann}}, \bibinfo {author} {\bibfnamefont {M.}~\bibnamefont {Loncar}},
  \emph {et~al.},\ }\href@noop {} {\bibfield  {journal} {\bibinfo  {journal}
  {Nature nanotechnology}\ }\textbf {\bibinfo {volume} {8}},\ \bibinfo {pages}
  {487} (\bibinfo {year} {2013})}\BibitemShut {NoStop}%
\bibitem [{\citenamefont {Neumann}\ \emph {et~al.}(2010)\citenamefont
  {Neumann}, \citenamefont {Beck}, \citenamefont {Steiner}, \citenamefont
  {Rempp}, \citenamefont {Fedder}, \citenamefont {Hemmer}, \citenamefont
  {Wrachtrup},\ and\ \citenamefont {Jelezko}}]{neumann2010single}%
  \BibitemOpen
  \bibfield  {author} {\bibinfo {author} {\bibfnamefont {P.}~\bibnamefont
  {Neumann}}, \bibinfo {author} {\bibfnamefont {J.}~\bibnamefont {Beck}},
  \bibinfo {author} {\bibfnamefont {M.}~\bibnamefont {Steiner}}, \bibinfo
  {author} {\bibfnamefont {F.}~\bibnamefont {Rempp}}, \bibinfo {author}
  {\bibfnamefont {H.}~\bibnamefont {Fedder}}, \bibinfo {author} {\bibfnamefont
  {P.~R.}\ \bibnamefont {Hemmer}}, \bibinfo {author} {\bibfnamefont
  {J.}~\bibnamefont {Wrachtrup}}, \ and\ \bibinfo {author} {\bibfnamefont
  {F.}~\bibnamefont {Jelezko}},\ }\href@noop {} {\bibfield  {journal} {\bibinfo
   {journal} {Science}\ }\textbf {\bibinfo {volume} {329}},\ \bibinfo {pages}
  {542} (\bibinfo {year} {2010})}\BibitemShut {NoStop}%
\bibitem [{\citenamefont {Pfender}\ \emph {et~al.}(2018)\citenamefont
  {Pfender}, \citenamefont {Wang}, \citenamefont {Sumiya}, \citenamefont
  {Onoda}, \citenamefont {Yang}, \citenamefont {Dasari}, \citenamefont
  {Neumann}, \citenamefont {Pan}, \citenamefont {Isoya}, \citenamefont {Liu}
  \emph {et~al.}}]{pfender2018back}%
  \BibitemOpen
  \bibfield  {author} {\bibinfo {author} {\bibfnamefont {M.}~\bibnamefont
  {Pfender}}, \bibinfo {author} {\bibfnamefont {P.}~\bibnamefont {Wang}},
  \bibinfo {author} {\bibfnamefont {H.}~\bibnamefont {Sumiya}}, \bibinfo
  {author} {\bibfnamefont {S.}~\bibnamefont {Onoda}}, \bibinfo {author}
  {\bibfnamefont {W.}~\bibnamefont {Yang}}, \bibinfo {author} {\bibfnamefont
  {D.~B.~R.}\ \bibnamefont {Dasari}}, \bibinfo {author} {\bibfnamefont
  {P.}~\bibnamefont {Neumann}}, \bibinfo {author} {\bibfnamefont {X.-Y.}\
  \bibnamefont {Pan}}, \bibinfo {author} {\bibfnamefont {J.}~\bibnamefont
  {Isoya}}, \bibinfo {author} {\bibfnamefont {R.-B.}\ \bibnamefont {Liu}},
  \emph {et~al.},\ }\href@noop {} {\bibfield  {journal} {\bibinfo  {journal}
  {arXiv preprint arXiv:1806.02181}\ } (\bibinfo {year} {2018})}\BibitemShut
  {NoStop}%
\bibitem [{\citenamefont {Unden}\ \emph {et~al.}(2016)\citenamefont {Unden},
  \citenamefont {Balasubramanian}, \citenamefont {Louzon}, \citenamefont
  {Vinkler}, \citenamefont {Plenio}, \citenamefont {Markham}, \citenamefont
  {Twitchen}, \citenamefont {Stacey}, \citenamefont {Lovchinsky}, \citenamefont
  {Sushkov}, \citenamefont {Lukin}, \citenamefont {Retzker}, \citenamefont
  {Naydenov}, \citenamefont {McGuinness},\ and\ \citenamefont
  {Jelezko}}]{unden2016quantum}%
  \BibitemOpen
  \bibfield  {author} {\bibinfo {author} {\bibfnamefont {T.}~\bibnamefont
  {Unden}}, \bibinfo {author} {\bibfnamefont {P.}~\bibnamefont
  {Balasubramanian}}, \bibinfo {author} {\bibfnamefont {D.}~\bibnamefont
  {Louzon}}, \bibinfo {author} {\bibfnamefont {Y.}~\bibnamefont {Vinkler}},
  \bibinfo {author} {\bibfnamefont {M.~B.}\ \bibnamefont {Plenio}}, \bibinfo
  {author} {\bibfnamefont {M.}~\bibnamefont {Markham}}, \bibinfo {author}
  {\bibfnamefont {D.}~\bibnamefont {Twitchen}}, \bibinfo {author}
  {\bibfnamefont {A.}~\bibnamefont {Stacey}}, \bibinfo {author} {\bibfnamefont
  {I.}~\bibnamefont {Lovchinsky}}, \bibinfo {author} {\bibfnamefont {A.~O.}\
  \bibnamefont {Sushkov}}, \bibinfo {author} {\bibfnamefont {M.}~\bibnamefont
  {Lukin}}, \bibinfo {author} {\bibfnamefont {A.}~\bibnamefont {Retzker}},
  \bibinfo {author} {\bibfnamefont {B.}~\bibnamefont {Naydenov}}, \bibinfo
  {author} {\bibfnamefont {L.~P.}\ \bibnamefont {McGuinness}}, \ and\ \bibinfo
  {author} {\bibfnamefont {F.}~\bibnamefont {Jelezko}},\ }\href@noop {}
  {\bibfield  {journal} {\bibinfo  {journal} {Physical Review Letters}\
  }\textbf {\bibinfo {volume} {116}},\ \bibinfo {pages} {230502} (\bibinfo
  {year} {2016})}\BibitemShut {NoStop}%
\bibitem [{\citenamefont {Jiang}\ \emph {et~al.}(2009)\citenamefont {Jiang},
  \citenamefont {Hodges}, \citenamefont {Maze}, \citenamefont {Maurer},
  \citenamefont {Taylor}, \citenamefont {Cory}, \citenamefont {Hemmer},
  \citenamefont {Walsworth}, \citenamefont {Yacoby}, \citenamefont {Zibrov}
  \emph {et~al.}}]{jiang2009repetitive}%
  \BibitemOpen
  \bibfield  {author} {\bibinfo {author} {\bibfnamefont {L.}~\bibnamefont
  {Jiang}}, \bibinfo {author} {\bibfnamefont {J.}~\bibnamefont {Hodges}},
  \bibinfo {author} {\bibfnamefont {J.}~\bibnamefont {Maze}}, \bibinfo {author}
  {\bibfnamefont {P.}~\bibnamefont {Maurer}}, \bibinfo {author} {\bibfnamefont
  {J.}~\bibnamefont {Taylor}}, \bibinfo {author} {\bibfnamefont
  {D.}~\bibnamefont {Cory}}, \bibinfo {author} {\bibfnamefont {P.}~\bibnamefont
  {Hemmer}}, \bibinfo {author} {\bibfnamefont {R.}~\bibnamefont {Walsworth}},
  \bibinfo {author} {\bibfnamefont {A.}~\bibnamefont {Yacoby}}, \bibinfo
  {author} {\bibfnamefont {A.}~\bibnamefont {Zibrov}},  \emph {et~al.},\
  }\href@noop {} {\bibfield  {journal} {\bibinfo  {journal} {Science}\ }\textbf
  {\bibinfo {volume} {326}},\ \bibinfo {pages} {267} (\bibinfo {year}
  {2009})}\BibitemShut {NoStop}%
\bibitem [{\citenamefont {Dutt}\ \emph {et~al.}(2007)\citenamefont {Dutt},
  \citenamefont {Childress}, \citenamefont {Jiang}, \citenamefont {Togan},
  \citenamefont {Maze}, \citenamefont {Jelezko}, \citenamefont {Zibrov},
  \citenamefont {Hemmer},\ and\ \citenamefont {Lukin}}]{dutt2007quantum}%
  \BibitemOpen
  \bibfield  {author} {\bibinfo {author} {\bibfnamefont {M.~G.}\ \bibnamefont
  {Dutt}}, \bibinfo {author} {\bibfnamefont {L.}~\bibnamefont {Childress}},
  \bibinfo {author} {\bibfnamefont {L.}~\bibnamefont {Jiang}}, \bibinfo
  {author} {\bibfnamefont {E.}~\bibnamefont {Togan}}, \bibinfo {author}
  {\bibfnamefont {J.}~\bibnamefont {Maze}}, \bibinfo {author} {\bibfnamefont
  {F.}~\bibnamefont {Jelezko}}, \bibinfo {author} {\bibfnamefont
  {A.}~\bibnamefont {Zibrov}}, \bibinfo {author} {\bibfnamefont
  {P.}~\bibnamefont {Hemmer}}, \ and\ \bibinfo {author} {\bibfnamefont
  {M.}~\bibnamefont {Lukin}},\ }\href@noop {} {\bibfield  {journal} {\bibinfo
  {journal} {Science}\ }\textbf {\bibinfo {volume} {316}},\ \bibinfo {pages}
  {1312} (\bibinfo {year} {2007})}\BibitemShut {NoStop}%
\bibitem [{\citenamefont {Falk}\ \emph {et~al.}(2015)\citenamefont {Falk},
  \citenamefont {Klimov}, \citenamefont {Iv{\'a}dy}, \citenamefont {Sz{\'a}sz},
  \citenamefont {Christle}, \citenamefont {Koehl}, \citenamefont {Gali},\ and\
  \citenamefont {Awschalom}}]{falk2015optical}%
  \BibitemOpen
  \bibfield  {author} {\bibinfo {author} {\bibfnamefont {A.~L.}\ \bibnamefont
  {Falk}}, \bibinfo {author} {\bibfnamefont {P.~V.}\ \bibnamefont {Klimov}},
  \bibinfo {author} {\bibfnamefont {V.}~\bibnamefont {Iv{\'a}dy}}, \bibinfo
  {author} {\bibfnamefont {K.}~\bibnamefont {Sz{\'a}sz}}, \bibinfo {author}
  {\bibfnamefont {D.~J.}\ \bibnamefont {Christle}}, \bibinfo {author}
  {\bibfnamefont {W.~F.}\ \bibnamefont {Koehl}}, \bibinfo {author}
  {\bibfnamefont {{\'A}.}~\bibnamefont {Gali}}, \ and\ \bibinfo {author}
  {\bibfnamefont {D.~D.}\ \bibnamefont {Awschalom}},\ }\href@noop {} {\bibfield
   {journal} {\bibinfo  {journal} {Physical review letters}\ }\textbf {\bibinfo
  {volume} {114}},\ \bibinfo {pages} {247603} (\bibinfo {year}
  {2015})}\BibitemShut {NoStop}%
\bibitem [{\citenamefont {Iv{\'a}dy}\ \emph {et~al.}(2015)\citenamefont
  {Iv{\'a}dy}, \citenamefont {Sz{\'a}sz}, \citenamefont {Falk}, \citenamefont
  {Klimov}, \citenamefont {Christle}, \citenamefont {Janz{\'e}n}, \citenamefont
  {Abrikosov}, \citenamefont {Awschalom},\ and\ \citenamefont
  {Gali}}]{ivady2015theoretical}%
  \BibitemOpen
  \bibfield  {author} {\bibinfo {author} {\bibfnamefont {V.}~\bibnamefont
  {Iv{\'a}dy}}, \bibinfo {author} {\bibfnamefont {K.}~\bibnamefont
  {Sz{\'a}sz}}, \bibinfo {author} {\bibfnamefont {A.~L.}\ \bibnamefont {Falk}},
  \bibinfo {author} {\bibfnamefont {P.~V.}\ \bibnamefont {Klimov}}, \bibinfo
  {author} {\bibfnamefont {D.~J.}\ \bibnamefont {Christle}}, \bibinfo {author}
  {\bibfnamefont {E.}~\bibnamefont {Janz{\'e}n}}, \bibinfo {author}
  {\bibfnamefont {I.~A.}\ \bibnamefont {Abrikosov}}, \bibinfo {author}
  {\bibfnamefont {D.~D.}\ \bibnamefont {Awschalom}}, \ and\ \bibinfo {author}
  {\bibfnamefont {A.}~\bibnamefont {Gali}},\ }\href@noop {} {\bibfield
  {journal} {\bibinfo  {journal} {Physical Review B}\ }\textbf {\bibinfo
  {volume} {92}},\ \bibinfo {pages} {115206} (\bibinfo {year}
  {2015})}\BibitemShut {NoStop}%
\bibitem [{\citenamefont {Morton}\ \emph {et~al.}(2008)\citenamefont {Morton},
  \citenamefont {Tyryshkin}, \citenamefont {Brown}, \citenamefont {Shankar},
  \citenamefont {Lovett}, \citenamefont {Ardavan}, \citenamefont {Schenkel},
  \citenamefont {Haller}, \citenamefont {Ager},\ and\ \citenamefont
  {Lyon}}]{morton2008solid}%
  \BibitemOpen
  \bibfield  {author} {\bibinfo {author} {\bibfnamefont {J.~J.}\ \bibnamefont
  {Morton}}, \bibinfo {author} {\bibfnamefont {A.~M.}\ \bibnamefont
  {Tyryshkin}}, \bibinfo {author} {\bibfnamefont {R.~M.}\ \bibnamefont
  {Brown}}, \bibinfo {author} {\bibfnamefont {S.}~\bibnamefont {Shankar}},
  \bibinfo {author} {\bibfnamefont {B.~W.}\ \bibnamefont {Lovett}}, \bibinfo
  {author} {\bibfnamefont {A.}~\bibnamefont {Ardavan}}, \bibinfo {author}
  {\bibfnamefont {T.}~\bibnamefont {Schenkel}}, \bibinfo {author}
  {\bibfnamefont {E.~E.}\ \bibnamefont {Haller}}, \bibinfo {author}
  {\bibfnamefont {J.~W.}\ \bibnamefont {Ager}}, \ and\ \bibinfo {author}
  {\bibfnamefont {S.}~\bibnamefont {Lyon}},\ }\href@noop {} {\bibfield
  {journal} {\bibinfo  {journal} {Nature}\ }\textbf {\bibinfo {volume} {455}},\
  \bibinfo {pages} {1085} (\bibinfo {year} {2008})}\BibitemShut {NoStop}%
\bibitem [{\citenamefont {Pla}\ \emph {et~al.}(2013)\citenamefont {Pla},
  \citenamefont {Tan}, \citenamefont {Dehollain}, \citenamefont {Lim},
  \citenamefont {Morton}, \citenamefont {Zwanenburg}, \citenamefont {Jamieson},
  \citenamefont {Dzurak},\ and\ \citenamefont {Morello}}]{pla2013high}%
  \BibitemOpen
  \bibfield  {author} {\bibinfo {author} {\bibfnamefont {J.~J.}\ \bibnamefont
  {Pla}}, \bibinfo {author} {\bibfnamefont {K.~Y.}\ \bibnamefont {Tan}},
  \bibinfo {author} {\bibfnamefont {J.~P.}\ \bibnamefont {Dehollain}}, \bibinfo
  {author} {\bibfnamefont {W.~H.}\ \bibnamefont {Lim}}, \bibinfo {author}
  {\bibfnamefont {J.~J.}\ \bibnamefont {Morton}}, \bibinfo {author}
  {\bibfnamefont {F.~A.}\ \bibnamefont {Zwanenburg}}, \bibinfo {author}
  {\bibfnamefont {D.~N.}\ \bibnamefont {Jamieson}}, \bibinfo {author}
  {\bibfnamefont {A.~S.}\ \bibnamefont {Dzurak}}, \ and\ \bibinfo {author}
  {\bibfnamefont {A.}~\bibnamefont {Morello}},\ }\href@noop {} {\bibfield
  {journal} {\bibinfo  {journal} {Nature}\ }\textbf {\bibinfo {volume} {496}},\
  \bibinfo {pages} {334} (\bibinfo {year} {2013})}\BibitemShut {NoStop}%
\bibitem [{\citenamefont {Yao}\ \emph {et~al.}(2012)\citenamefont {Yao},
  \citenamefont {Jiang}, \citenamefont {Gorshkov}, \citenamefont {Maurer},
  \citenamefont {Giedke}, \citenamefont {Cirac},\ and\ \citenamefont
  {Lukin}}]{yao2012scalable}%
  \BibitemOpen
  \bibfield  {author} {\bibinfo {author} {\bibfnamefont {N.~Y.}\ \bibnamefont
  {Yao}}, \bibinfo {author} {\bibfnamefont {L.}~\bibnamefont {Jiang}}, \bibinfo
  {author} {\bibfnamefont {A.~V.}\ \bibnamefont {Gorshkov}}, \bibinfo {author}
  {\bibfnamefont {P.~C.}\ \bibnamefont {Maurer}}, \bibinfo {author}
  {\bibfnamefont {G.}~\bibnamefont {Giedke}}, \bibinfo {author} {\bibfnamefont
  {J.~I.}\ \bibnamefont {Cirac}}, \ and\ \bibinfo {author} {\bibfnamefont
  {M.~D.}\ \bibnamefont {Lukin}},\ }\href@noop {} {\bibfield  {journal}
  {\bibinfo  {journal} {Nature communications}\ }\textbf {\bibinfo {volume}
  {3}},\ \bibinfo {pages} {800} (\bibinfo {year} {2012})}\BibitemShut {NoStop}%
\bibitem [{\citenamefont {Childress}\ and\ \citenamefont
  {Hanson}(2013)}]{childress2013diamond}%
  \BibitemOpen
  \bibfield  {author} {\bibinfo {author} {\bibfnamefont {L.}~\bibnamefont
  {Childress}}\ and\ \bibinfo {author} {\bibfnamefont {R.}~\bibnamefont
  {Hanson}},\ }\href@noop {} {\bibfield  {journal} {\bibinfo  {journal} {MRS
  bulletin}\ }\textbf {\bibinfo {volume} {38}},\ \bibinfo {pages} {134}
  (\bibinfo {year} {2013})}\BibitemShut {NoStop}%
\bibitem [{\citenamefont {Robledo}\ \emph {et~al.}(2011)\citenamefont
  {Robledo}, \citenamefont {Childress}, \citenamefont {Bernien}, \citenamefont
  {Hensen}, \citenamefont {Alkemade},\ and\ \citenamefont
  {Hanson}}]{robledo2011high}%
  \BibitemOpen
  \bibfield  {author} {\bibinfo {author} {\bibfnamefont {L.}~\bibnamefont
  {Robledo}}, \bibinfo {author} {\bibfnamefont {L.}~\bibnamefont {Childress}},
  \bibinfo {author} {\bibfnamefont {H.}~\bibnamefont {Bernien}}, \bibinfo
  {author} {\bibfnamefont {B.}~\bibnamefont {Hensen}}, \bibinfo {author}
  {\bibfnamefont {P.~F.}\ \bibnamefont {Alkemade}}, \ and\ \bibinfo {author}
  {\bibfnamefont {R.}~\bibnamefont {Hanson}},\ }\href@noop {} {\bibfield
  {journal} {\bibinfo  {journal} {Nature}\ }\textbf {\bibinfo {volume} {477}},\
  \bibinfo {pages} {574} (\bibinfo {year} {2011})}\BibitemShut {NoStop}%
\bibitem [{\citenamefont {Van~der Sar}\ \emph {et~al.}(2012)\citenamefont
  {Van~der Sar}, \citenamefont {Wang}, \citenamefont {Blok}, \citenamefont
  {Bernien}, \citenamefont {Taminiau}, \citenamefont {Toyli}, \citenamefont
  {Lidar}, \citenamefont {Awschalom}, \citenamefont {Hanson},\ and\
  \citenamefont {Dobrovitski}}]{van2012decoherence}%
  \BibitemOpen
  \bibfield  {author} {\bibinfo {author} {\bibfnamefont {T.}~\bibnamefont
  {Van~der Sar}}, \bibinfo {author} {\bibfnamefont {Z.}~\bibnamefont {Wang}},
  \bibinfo {author} {\bibfnamefont {M.}~\bibnamefont {Blok}}, \bibinfo {author}
  {\bibfnamefont {H.}~\bibnamefont {Bernien}}, \bibinfo {author} {\bibfnamefont
  {T.}~\bibnamefont {Taminiau}}, \bibinfo {author} {\bibfnamefont
  {D.}~\bibnamefont {Toyli}}, \bibinfo {author} {\bibfnamefont
  {D.}~\bibnamefont {Lidar}}, \bibinfo {author} {\bibfnamefont
  {D.}~\bibnamefont {Awschalom}}, \bibinfo {author} {\bibfnamefont
  {R.}~\bibnamefont {Hanson}}, \ and\ \bibinfo {author} {\bibfnamefont
  {V.}~\bibnamefont {Dobrovitski}},\ }\href@noop {} {\bibfield  {journal}
  {\bibinfo  {journal} {Nature}\ }\textbf {\bibinfo {volume} {484}},\ \bibinfo
  {pages} {82} (\bibinfo {year} {2012})}\BibitemShut {NoStop}%
\bibitem [{\citenamefont {Taminiau}\ \emph {et~al.}(2014)\citenamefont
  {Taminiau}, \citenamefont {Cramer}, \citenamefont {van~der Sar},
  \citenamefont {Dobrovitski},\ and\ \citenamefont
  {Hanson}}]{taminiau2014universal}%
  \BibitemOpen
  \bibfield  {author} {\bibinfo {author} {\bibfnamefont {T.~H.}\ \bibnamefont
  {Taminiau}}, \bibinfo {author} {\bibfnamefont {J.}~\bibnamefont {Cramer}},
  \bibinfo {author} {\bibfnamefont {T.}~\bibnamefont {van~der Sar}}, \bibinfo
  {author} {\bibfnamefont {V.~V.}\ \bibnamefont {Dobrovitski}}, \ and\ \bibinfo
  {author} {\bibfnamefont {R.}~\bibnamefont {Hanson}},\ }\href@noop {}
  {\bibfield  {journal} {\bibinfo  {journal} {Nature nanotechnology}\ }\textbf
  {\bibinfo {volume} {9}},\ \bibinfo {pages} {171} (\bibinfo {year}
  {2014})}\BibitemShut {NoStop}%
\bibitem [{\citenamefont {Aslam}\ \emph {et~al.}(2018)\citenamefont {Aslam},
  \citenamefont {Pfender}, \citenamefont {Neumann}, \citenamefont {Reuter},
  \citenamefont {Zappe}, \citenamefont {F{\'a}varo~de Oliveira}, \citenamefont
  {Denisenko}, \citenamefont {Sumiya}, \citenamefont {Onada}, \citenamefont
  {Isoya} \emph {et~al.}}]{aslam2018nanoscale}%
  \BibitemOpen
  \bibfield  {author} {\bibinfo {author} {\bibfnamefont {N.}~\bibnamefont
  {Aslam}}, \bibinfo {author} {\bibfnamefont {M.}~\bibnamefont {Pfender}},
  \bibinfo {author} {\bibfnamefont {P.}~\bibnamefont {Neumann}}, \bibinfo
  {author} {\bibfnamefont {R.}~\bibnamefont {Reuter}}, \bibinfo {author}
  {\bibfnamefont {A.}~\bibnamefont {Zappe}}, \bibinfo {author} {\bibfnamefont
  {F.}~\bibnamefont {F{\'a}varo~de Oliveira}}, \bibinfo {author} {\bibfnamefont
  {A.}~\bibnamefont {Denisenko}}, \bibinfo {author} {\bibfnamefont
  {H.}~\bibnamefont {Sumiya}}, \bibinfo {author} {\bibfnamefont
  {S.}~\bibnamefont {Onada}}, \bibinfo {author} {\bibfnamefont
  {J.}~\bibnamefont {Isoya}},  \emph {et~al.},\ }\href@noop {} {\bibfield
  {journal} {\bibinfo  {journal} {Bulletin of the American Physical Society}\ }
  (\bibinfo {year} {2018})}\BibitemShut {NoStop}%
\bibitem [{\citenamefont {Perunicic}\ \emph {et~al.}(2014)\citenamefont
  {Perunicic}, \citenamefont {Hall}, \citenamefont {Simpson}, \citenamefont
  {Hill},\ and\ \citenamefont {Hollenberg}}]{perunicic2014towards}%
  \BibitemOpen
  \bibfield  {author} {\bibinfo {author} {\bibfnamefont {V.~S.}\ \bibnamefont
  {Perunicic}}, \bibinfo {author} {\bibfnamefont {L.~T.}\ \bibnamefont {Hall}},
  \bibinfo {author} {\bibfnamefont {D.~A.}\ \bibnamefont {Simpson}}, \bibinfo
  {author} {\bibfnamefont {C.~D.}\ \bibnamefont {Hill}}, \ and\ \bibinfo
  {author} {\bibfnamefont {L.~C.}\ \bibnamefont {Hollenberg}},\ }\href@noop {}
  {\bibfield  {journal} {\bibinfo  {journal} {Physical Review B}\ }\textbf
  {\bibinfo {volume} {89}},\ \bibinfo {pages} {054432} (\bibinfo {year}
  {2014})}\BibitemShut {NoStop}%
\bibitem [{\citenamefont {Kong}\ \emph {et~al.}(2017)\citenamefont {Kong},
  \citenamefont {Shi}, \citenamefont {Yang}, \citenamefont {Wang},
  \citenamefont {Raatz}, \citenamefont {Meijer},\ and\ \citenamefont
  {Du}}]{kong2017atomic}%
  \BibitemOpen
  \bibfield  {author} {\bibinfo {author} {\bibfnamefont {X.}~\bibnamefont
  {Kong}}, \bibinfo {author} {\bibfnamefont {F.}~\bibnamefont {Shi}}, \bibinfo
  {author} {\bibfnamefont {Z.}~\bibnamefont {Yang}}, \bibinfo {author}
  {\bibfnamefont {P.}~\bibnamefont {Wang}}, \bibinfo {author} {\bibfnamefont
  {N.}~\bibnamefont {Raatz}}, \bibinfo {author} {\bibfnamefont
  {J.}~\bibnamefont {Meijer}}, \ and\ \bibinfo {author} {\bibfnamefont
  {J.}~\bibnamefont {Du}},\ }\href@noop {} {\bibfield  {journal} {\bibinfo
  {journal} {arXiv preprint arXiv:1705.09201}\ } (\bibinfo {year}
  {2017})}\BibitemShut {NoStop}%
\bibitem [{\citenamefont {Shagieva}\ \emph {et~al.}(2018)\citenamefont
  {Shagieva}, \citenamefont {Zaiser}, \citenamefont {Neumann}, \citenamefont
  {Dasari}, \citenamefont {St�hr}, \citenamefont {Denisenko}, \citenamefont
  {Reuter}, \citenamefont {Meriles},\ and\ \citenamefont
  {Wrachtrup}}]{shagieva2018microwave}%
  \BibitemOpen
  \bibfield  {author} {\bibinfo {author} {\bibfnamefont {F.}~\bibnamefont
  {Shagieva}}, \bibinfo {author} {\bibfnamefont {S.}~\bibnamefont {Zaiser}},
  \bibinfo {author} {\bibfnamefont {P.}~\bibnamefont {Neumann}}, \bibinfo
  {author} {\bibfnamefont {D.~B.~R.}\ \bibnamefont {Dasari}}, \bibinfo {author}
  {\bibfnamefont {R.}~\bibnamefont {St�hr}}, \bibinfo {author} {\bibfnamefont
  {A.}~\bibnamefont {Denisenko}}, \bibinfo {author} {\bibfnamefont
  {R.}~\bibnamefont {Reuter}}, \bibinfo {author} {\bibfnamefont {C.~A.}\
  \bibnamefont {Meriles}}, \ and\ \bibinfo {author} {\bibfnamefont
  {J.}~\bibnamefont {Wrachtrup}},\ }\href@noop {} {\bibfield  {journal}
  {\bibinfo  {journal} {Nano letters}\ }\textbf {\bibinfo {volume} {18}},\
  \bibinfo {pages} {3731} (\bibinfo {year} {2018})}\BibitemShut {NoStop}%
\bibitem [{\citenamefont {Pagliero}\ \emph {et~al.}(2018)\citenamefont
  {Pagliero}, \citenamefont {Rao}, \citenamefont {Zangara}, \citenamefont
  {Dhomkar}, \citenamefont {Wong}, \citenamefont {Abril}, \citenamefont
  {Aslam}, \citenamefont {Parker}, \citenamefont {King}, \citenamefont {Avalos}
  \emph {et~al.}}]{pagliero2018multispin}%
  \BibitemOpen
  \bibfield  {author} {\bibinfo {author} {\bibfnamefont {D.}~\bibnamefont
  {Pagliero}}, \bibinfo {author} {\bibfnamefont {K.~K.}\ \bibnamefont {Rao}},
  \bibinfo {author} {\bibfnamefont {P.~R.}\ \bibnamefont {Zangara}}, \bibinfo
  {author} {\bibfnamefont {S.}~\bibnamefont {Dhomkar}}, \bibinfo {author}
  {\bibfnamefont {H.~H.}\ \bibnamefont {Wong}}, \bibinfo {author}
  {\bibfnamefont {A.}~\bibnamefont {Abril}}, \bibinfo {author} {\bibfnamefont
  {N.}~\bibnamefont {Aslam}}, \bibinfo {author} {\bibfnamefont
  {A.}~\bibnamefont {Parker}}, \bibinfo {author} {\bibfnamefont
  {J.}~\bibnamefont {King}}, \bibinfo {author} {\bibfnamefont {C.~E.}\
  \bibnamefont {Avalos}},  \emph {et~al.},\ }\href@noop {} {\bibfield
  {journal} {\bibinfo  {journal} {Physical Review B}\ }\textbf {\bibinfo
  {volume} {97}},\ \bibinfo {pages} {024422} (\bibinfo {year}
  {2018})}\BibitemShut {NoStop}%
\bibitem [{\citenamefont {Broadway}\ \emph {et~al.}(2018)\citenamefont
  {Broadway}, \citenamefont {Tetienne}, \citenamefont {Stacey}, \citenamefont
  {Wood}, \citenamefont {Simpson}, \citenamefont {Hall},\ and\ \citenamefont
  {Hollenberg}}]{broadway2018quantum}%
  \BibitemOpen
  \bibfield  {author} {\bibinfo {author} {\bibfnamefont {D.~A.}\ \bibnamefont
  {Broadway}}, \bibinfo {author} {\bibfnamefont {J.-P.}\ \bibnamefont
  {Tetienne}}, \bibinfo {author} {\bibfnamefont {A.}~\bibnamefont {Stacey}},
  \bibinfo {author} {\bibfnamefont {J.~D.}\ \bibnamefont {Wood}}, \bibinfo
  {author} {\bibfnamefont {D.~A.}\ \bibnamefont {Simpson}}, \bibinfo {author}
  {\bibfnamefont {L.~T.}\ \bibnamefont {Hall}}, \ and\ \bibinfo {author}
  {\bibfnamefont {L.~C.}\ \bibnamefont {Hollenberg}},\ }\href@noop {}
  {\bibfield  {journal} {\bibinfo  {journal} {Nature communications}\ }\textbf
  {\bibinfo {volume} {9}},\ \bibinfo {pages} {1246} (\bibinfo {year}
  {2018})}\BibitemShut {NoStop}%
\bibitem [{\citenamefont {Fern{\'a}ndez-Acebal}\ \emph
  {et~al.}(2018)\citenamefont {Fern{\'a}ndez-Acebal}, \citenamefont {Rosolio},
  \citenamefont {Scheuer}, \citenamefont {M�ller}, \citenamefont {M�ller},
  \citenamefont {Schmitt}, \citenamefont {McGuinness}, \citenamefont {Schwarz},
  \citenamefont {Chen}, \citenamefont {Retzker} \emph
  {et~al.}}]{fernandez2018toward}%
  \BibitemOpen
  \bibfield  {author} {\bibinfo {author} {\bibfnamefont {P.}~\bibnamefont
  {Fern{\'a}ndez-Acebal}}, \bibinfo {author} {\bibfnamefont {O.}~\bibnamefont
  {Rosolio}}, \bibinfo {author} {\bibfnamefont {J.}~\bibnamefont {Scheuer}},
  \bibinfo {author} {\bibfnamefont {C.}~\bibnamefont {M�ller}}, \bibinfo
  {author} {\bibfnamefont {S.}~\bibnamefont {M�ller}}, \bibinfo {author}
  {\bibfnamefont {S.}~\bibnamefont {Schmitt}}, \bibinfo {author} {\bibfnamefont
  {L.}~\bibnamefont {McGuinness}}, \bibinfo {author} {\bibfnamefont
  {I.}~\bibnamefont {Schwarz}}, \bibinfo {author} {\bibfnamefont
  {Q.}~\bibnamefont {Chen}}, \bibinfo {author} {\bibfnamefont {A.}~\bibnamefont
  {Retzker}},  \emph {et~al.},\ }\href@noop {} {\bibfield  {journal} {\bibinfo
  {journal} {Nano letters}\ }\textbf {\bibinfo {volume} {18}},\ \bibinfo
  {pages} {1882} (\bibinfo {year} {2018})}\BibitemShut {NoStop}%
\bibitem [{\citenamefont {Scheuer}\ \emph {et~al.}(2016)\citenamefont
  {Scheuer}, \citenamefont {Schwartz}, \citenamefont {Chen}, \citenamefont
  {Schulze-S{\"u}nninghausen}, \citenamefont {Carl}, \citenamefont {H{\"o}fer},
  \citenamefont {Retzker}, \citenamefont {Sumiya}, \citenamefont {Isoya},
  \citenamefont {Luy} \emph {et~al.}}]{scheuer2016optically}%
  \BibitemOpen
  \bibfield  {author} {\bibinfo {author} {\bibfnamefont {J.}~\bibnamefont
  {Scheuer}}, \bibinfo {author} {\bibfnamefont {I.}~\bibnamefont {Schwartz}},
  \bibinfo {author} {\bibfnamefont {Q.}~\bibnamefont {Chen}}, \bibinfo {author}
  {\bibfnamefont {D.}~\bibnamefont {Schulze-S{\"u}nninghausen}}, \bibinfo
  {author} {\bibfnamefont {P.}~\bibnamefont {Carl}}, \bibinfo {author}
  {\bibfnamefont {P.}~\bibnamefont {H{\"o}fer}}, \bibinfo {author}
  {\bibfnamefont {A.}~\bibnamefont {Retzker}}, \bibinfo {author} {\bibfnamefont
  {H.}~\bibnamefont {Sumiya}}, \bibinfo {author} {\bibfnamefont
  {J.}~\bibnamefont {Isoya}}, \bibinfo {author} {\bibfnamefont
  {B.}~\bibnamefont {Luy}},  \emph {et~al.},\ }\href@noop {} {\bibfield
  {journal} {\bibinfo  {journal} {New Journal of Physics}\ }\textbf {\bibinfo
  {volume} {18}},\ \bibinfo {pages} {013040} (\bibinfo {year}
  {2016})}\BibitemShut {NoStop}%
\bibitem [{\citenamefont {Chen}\ \emph {et~al.}(2016)\citenamefont {Chen},
  \citenamefont {Schwarz}, \citenamefont {Jelezko}, \citenamefont {Retzker},\
  and\ \citenamefont {Plenio}}]{chen2016resonance}%
  \BibitemOpen
  \bibfield  {author} {\bibinfo {author} {\bibfnamefont {Q.}~\bibnamefont
  {Chen}}, \bibinfo {author} {\bibfnamefont {I.}~\bibnamefont {Schwarz}},
  \bibinfo {author} {\bibfnamefont {F.}~\bibnamefont {Jelezko}}, \bibinfo
  {author} {\bibfnamefont {A.}~\bibnamefont {Retzker}}, \ and\ \bibinfo
  {author} {\bibfnamefont {M.~B.}\ \bibnamefont {Plenio}},\ }\href@noop {}
  {\bibfield  {journal} {\bibinfo  {journal} {Physical Review B}\ }\textbf
  {\bibinfo {volume} {93}},\ \bibinfo {pages} {060408} (\bibinfo {year}
  {2016})}\BibitemShut {NoStop}%
\bibitem [{\citenamefont {Hartmann}\ and\ \citenamefont
  {Hahn}(1962)}]{hartmann1962nuclear}%
  \BibitemOpen
  \bibfield  {author} {\bibinfo {author} {\bibfnamefont {S.}~\bibnamefont
  {Hartmann}}\ and\ \bibinfo {author} {\bibfnamefont {E.}~\bibnamefont
  {Hahn}},\ }\href@noop {} {\bibfield  {journal} {\bibinfo  {journal} {Physical
  Review}\ }\textbf {\bibinfo {volume} {128}},\ \bibinfo {pages} {2042}
  (\bibinfo {year} {1962})}\BibitemShut {NoStop}%
\bibitem [{\citenamefont {Aslam}\ \emph {et~al.}(2015)\citenamefont {Aslam},
  \citenamefont {Pfender}, \citenamefont {St{\"o}hr}, \citenamefont {Neumann},
  \citenamefont {Scheffler}, \citenamefont {Sumiya}, \citenamefont {Abe},
  \citenamefont {Onoda}, \citenamefont {Ohshima}, \citenamefont {Isoya} \emph
  {et~al.}}]{aslam2015single}%
  \BibitemOpen
  \bibfield  {author} {\bibinfo {author} {\bibfnamefont {N.}~\bibnamefont
  {Aslam}}, \bibinfo {author} {\bibfnamefont {M.}~\bibnamefont {Pfender}},
  \bibinfo {author} {\bibfnamefont {R.}~\bibnamefont {St{\"o}hr}}, \bibinfo
  {author} {\bibfnamefont {P.}~\bibnamefont {Neumann}}, \bibinfo {author}
  {\bibfnamefont {M.}~\bibnamefont {Scheffler}}, \bibinfo {author}
  {\bibfnamefont {H.}~\bibnamefont {Sumiya}}, \bibinfo {author} {\bibfnamefont
  {H.}~\bibnamefont {Abe}}, \bibinfo {author} {\bibfnamefont {S.}~\bibnamefont
  {Onoda}}, \bibinfo {author} {\bibfnamefont {T.}~\bibnamefont {Ohshima}},
  \bibinfo {author} {\bibfnamefont {J.}~\bibnamefont {Isoya}},  \emph
  {et~al.},\ }\href@noop {} {\bibfield  {journal} {\bibinfo  {journal} {Review
  of Scientific Instruments}\ }\textbf {\bibinfo {volume} {86}},\ \bibinfo
  {pages} {064704} (\bibinfo {year} {2015})}\BibitemShut {NoStop}%
\bibitem [{\citenamefont {H{\"a}berle}\ \emph {et~al.}(2017)\citenamefont
  {H{\"a}berle}, \citenamefont {Oeckinghaus}, \citenamefont {Schmid-Lorch},
  \citenamefont {Pfender}, \citenamefont {de~Oliveira}, \citenamefont
  {Momenzadeh}, \citenamefont {Finkler},\ and\ \citenamefont
  {Wrachtrup}}]{haberle2017nuclear}%
  \BibitemOpen
  \bibfield  {author} {\bibinfo {author} {\bibfnamefont {T.}~\bibnamefont
  {H{\"a}berle}}, \bibinfo {author} {\bibfnamefont {T.}~\bibnamefont
  {Oeckinghaus}}, \bibinfo {author} {\bibfnamefont {D.}~\bibnamefont
  {Schmid-Lorch}}, \bibinfo {author} {\bibfnamefont {M.}~\bibnamefont
  {Pfender}}, \bibinfo {author} {\bibfnamefont {F.~F.}\ \bibnamefont
  {de~Oliveira}}, \bibinfo {author} {\bibfnamefont {S.~A.}\ \bibnamefont
  {Momenzadeh}}, \bibinfo {author} {\bibfnamefont {A.}~\bibnamefont {Finkler}},
  \ and\ \bibinfo {author} {\bibfnamefont {J.}~\bibnamefont {Wrachtrup}},\
  }\href@noop {} {\bibfield  {journal} {\bibinfo  {journal} {Review of
  Scientific Instruments}\ }\textbf {\bibinfo {volume} {88}},\ \bibinfo {pages}
  {013702} (\bibinfo {year} {2017})}\BibitemShut {NoStop}%
\bibitem [{\citenamefont {Stepanov}\ \emph {et~al.}(2015)\citenamefont
  {Stepanov}, \citenamefont {Cho}, \citenamefont {Abeywardana},\ and\
  \citenamefont {Takahashi}}]{stepanov2015high}%
  \BibitemOpen
  \bibfield  {author} {\bibinfo {author} {\bibfnamefont {V.}~\bibnamefont
  {Stepanov}}, \bibinfo {author} {\bibfnamefont {F.~H.}\ \bibnamefont {Cho}},
  \bibinfo {author} {\bibfnamefont {C.}~\bibnamefont {Abeywardana}}, \ and\
  \bibinfo {author} {\bibfnamefont {S.}~\bibnamefont {Takahashi}},\ }\href@noop
  {} {\bibfield  {journal} {\bibinfo  {journal} {Applied Physics Letters}\
  }\textbf {\bibinfo {volume} {106}},\ \bibinfo {pages} {063111} (\bibinfo
  {year} {2015})}\BibitemShut {NoStop}%
\bibitem [{\citenamefont {Pfender}\ \emph {et~al.}(2017)\citenamefont
  {Pfender}, \citenamefont {Aslam}, \citenamefont {Simon}, \citenamefont
  {Antonov}, \citenamefont {Thiering}, \citenamefont {Burk}, \citenamefont
  {F�varo~de Oliveira}, \citenamefont {Denisenko}, \citenamefont {Fedder},
  \citenamefont {Meijer} \emph {et~al.}}]{pfender2017protecting}%
  \BibitemOpen
  \bibfield  {author} {\bibinfo {author} {\bibfnamefont {M.}~\bibnamefont
  {Pfender}}, \bibinfo {author} {\bibfnamefont {N.}~\bibnamefont {Aslam}},
  \bibinfo {author} {\bibfnamefont {P.}~\bibnamefont {Simon}}, \bibinfo
  {author} {\bibfnamefont {D.}~\bibnamefont {Antonov}}, \bibinfo {author}
  {\bibfnamefont {G.}~\bibnamefont {Thiering}}, \bibinfo {author}
  {\bibfnamefont {S.}~\bibnamefont {Burk}}, \bibinfo {author} {\bibfnamefont
  {F.}~\bibnamefont {F�varo~de Oliveira}}, \bibinfo {author} {\bibfnamefont
  {A.}~\bibnamefont {Denisenko}}, \bibinfo {author} {\bibfnamefont
  {H.}~\bibnamefont {Fedder}}, \bibinfo {author} {\bibfnamefont
  {J.}~\bibnamefont {Meijer}},  \emph {et~al.},\ }\href@noop {} {\bibfield
  {journal} {\bibinfo  {journal} {Nano letters}\ }\textbf {\bibinfo {volume}
  {17}},\ \bibinfo {pages} {5931} (\bibinfo {year} {2017})}\BibitemShut
  {NoStop}%
\bibitem [{\citenamefont {Casanova}\ \emph {et~al.}(2018)\citenamefont
  {Casanova}, \citenamefont {Wang}, \citenamefont {Schwartz},\ and\
  \citenamefont {Plenio}}]{casanova2018shaped}%
  \BibitemOpen
  \bibfield  {author} {\bibinfo {author} {\bibfnamefont {J.}~\bibnamefont
  {Casanova}}, \bibinfo {author} {\bibfnamefont {Z.-Y.}\ \bibnamefont {Wang}},
  \bibinfo {author} {\bibfnamefont {I.}~\bibnamefont {Schwartz}}, \ and\
  \bibinfo {author} {\bibfnamefont {M.}~\bibnamefont {Plenio}},\ }\href@noop {}
  {\bibfield  {journal} {\bibinfo  {journal} {arXiv preprint arXiv:1805.01741}\
  } (\bibinfo {year} {2018})}\BibitemShut {NoStop}%
\bibitem [{\citenamefont {Gordon}\ \emph {et~al.}(2008)\citenamefont {Gordon},
  \citenamefont {Kurizki},\ and\ \citenamefont {Lidar}}]{gordon2008optimal}%
  \BibitemOpen
  \bibfield  {author} {\bibinfo {author} {\bibfnamefont {G.}~\bibnamefont
  {Gordon}}, \bibinfo {author} {\bibfnamefont {G.}~\bibnamefont {Kurizki}}, \
  and\ \bibinfo {author} {\bibfnamefont {D.~A.}\ \bibnamefont {Lidar}},\
  }\href@noop {} {\bibfield  {journal} {\bibinfo  {journal} {Physical review
  letters}\ }\textbf {\bibinfo {volume} {101}},\ \bibinfo {pages} {010403}
  (\bibinfo {year} {2008})}\BibitemShut {NoStop}%
\bibitem [{\citenamefont {Cao}\ \emph {et~al.}(2017)\citenamefont {Cao},
  \citenamefont {Shu}, \citenamefont {Yang}, \citenamefont {Yu}, \citenamefont
  {Gong}, \citenamefont {He}, \citenamefont {Hu}, \citenamefont {Retzker},
  \citenamefont {Plenio}, \citenamefont {M{\"u}ller} \emph
  {et~al.}}]{cao2017protecting}%
  \BibitemOpen
  \bibfield  {author} {\bibinfo {author} {\bibfnamefont {Q.-Y.}\ \bibnamefont
  {Cao}}, \bibinfo {author} {\bibfnamefont {Z.-J.}\ \bibnamefont {Shu}},
  \bibinfo {author} {\bibfnamefont {P.-C.}\ \bibnamefont {Yang}}, \bibinfo
  {author} {\bibfnamefont {M.}~\bibnamefont {Yu}}, \bibinfo {author}
  {\bibfnamefont {M.-S.}\ \bibnamefont {Gong}}, \bibinfo {author}
  {\bibfnamefont {J.-Y.}\ \bibnamefont {He}}, \bibinfo {author} {\bibfnamefont
  {R.-F.}\ \bibnamefont {Hu}}, \bibinfo {author} {\bibfnamefont
  {A.}~\bibnamefont {Retzker}}, \bibinfo {author} {\bibfnamefont
  {M.}~\bibnamefont {Plenio}}, \bibinfo {author} {\bibfnamefont
  {C.}~\bibnamefont {M{\"u}ller}},  \emph {et~al.},\ }\href@noop {} {\bibfield
  {journal} {\bibinfo  {journal} {arXiv preprint arXiv:1710.10744}\ } (\bibinfo
  {year} {2017})}\BibitemShut {NoStop}%
\bibitem [{sup()}]{supp}%
  \BibitemOpen
  \href@noop {} {\bibinfo  {journal} {See Supplemental Material for further
  details, which includes Refs.
  \cite{cohen1986quantum,gillespie1996exact,aharon2016fully}}\ }\BibitemShut
  {NoStop}%
\bibitem [{\citenamefont {Cohen}\ \emph {et~al.}(2017)\citenamefont {Cohen},
  \citenamefont {Aharon},\ and\ \citenamefont {Retzker}}]{cohen2017continuous}%
  \BibitemOpen
\bibfield  {journal} {  }\bibfield  {author} {\bibinfo {author} {\bibfnamefont
  {I.}~\bibnamefont {Cohen}}, \bibinfo {author} {\bibfnamefont
  {N.}~\bibnamefont {Aharon}}, \ and\ \bibinfo {author} {\bibfnamefont
  {A.}~\bibnamefont {Retzker}},\ }\href@noop {} {\bibfield  {journal} {\bibinfo
   {journal} {Fortschritte der Physik}\ }\textbf {\bibinfo {volume} {65}},\
  \bibinfo {pages} {1600071} (\bibinfo {year} {2017})}\BibitemShut {NoStop}%
\bibitem [{\citenamefont {Farfurnik}\ \emph {et~al.}(2017)\citenamefont
  {Farfurnik}, \citenamefont {Aharon}, \citenamefont {Cohen}, \citenamefont
  {Hovav}, \citenamefont {Retzker},\ and\ \citenamefont
  {Bar-Gill}}]{farfurnik2017experimental}%
  \BibitemOpen
  \bibfield  {author} {\bibinfo {author} {\bibfnamefont {D.}~\bibnamefont
  {Farfurnik}}, \bibinfo {author} {\bibfnamefont {N.}~\bibnamefont {Aharon}},
  \bibinfo {author} {\bibfnamefont {I.}~\bibnamefont {Cohen}}, \bibinfo
  {author} {\bibfnamefont {Y.}~\bibnamefont {Hovav}}, \bibinfo {author}
  {\bibfnamefont {A.}~\bibnamefont {Retzker}}, \ and\ \bibinfo {author}
  {\bibfnamefont {N.}~\bibnamefont {Bar-Gill}},\ }\href@noop {} {\bibfield
  {journal} {\bibinfo  {journal} {Physical Review A}\ }\textbf {\bibinfo
  {volume} {96}},\ \bibinfo {pages} {013850} (\bibinfo {year}
  {2017})}\BibitemShut {NoStop}%
\bibitem [{\citenamefont {Schmitt}\ \emph {et~al.}(2017)\citenamefont
  {Schmitt}, \citenamefont {Gefen}, \citenamefont {St{\"u}rner}, \citenamefont
  {Unden}, \citenamefont {Wolff}, \citenamefont {M{\"u}ller}, \citenamefont
  {Scheuer}, \citenamefont {Naydenov}, \citenamefont {Markham}, \citenamefont
  {Pezzagna} \emph {et~al.}}]{schmitt2017submillihertz}%
  \BibitemOpen
  \bibfield  {author} {\bibinfo {author} {\bibfnamefont {S.}~\bibnamefont
  {Schmitt}}, \bibinfo {author} {\bibfnamefont {T.}~\bibnamefont {Gefen}},
  \bibinfo {author} {\bibfnamefont {F.~M.}\ \bibnamefont {St{\"u}rner}},
  \bibinfo {author} {\bibfnamefont {T.}~\bibnamefont {Unden}}, \bibinfo
  {author} {\bibfnamefont {G.}~\bibnamefont {Wolff}}, \bibinfo {author}
  {\bibfnamefont {C.}~\bibnamefont {M{\"u}ller}}, \bibinfo {author}
  {\bibfnamefont {J.}~\bibnamefont {Scheuer}}, \bibinfo {author} {\bibfnamefont
  {B.}~\bibnamefont {Naydenov}}, \bibinfo {author} {\bibfnamefont
  {M.}~\bibnamefont {Markham}}, \bibinfo {author} {\bibfnamefont
  {S.}~\bibnamefont {Pezzagna}},  \emph {et~al.},\ }\href@noop {} {\bibfield
  {journal} {\bibinfo  {journal} {Science}\ }\textbf {\bibinfo {volume}
  {356}},\ \bibinfo {pages} {832} (\bibinfo {year} {2017})}\BibitemShut
  {NoStop}%
\bibitem [{\citenamefont {Bucher}\ \emph {et~al.}(2017)\citenamefont {Bucher},
  \citenamefont {Glenn}, \citenamefont {Lee}, \citenamefont {Lukin},
  \citenamefont {Park},\ and\ \citenamefont {Walsworth}}]{bucher2017high}%
  \BibitemOpen
  \bibfield  {author} {\bibinfo {author} {\bibfnamefont {D.~B.}\ \bibnamefont
  {Bucher}}, \bibinfo {author} {\bibfnamefont {D.~R.}\ \bibnamefont {Glenn}},
  \bibinfo {author} {\bibfnamefont {J.}~\bibnamefont {Lee}}, \bibinfo {author}
  {\bibfnamefont {M.~D.}\ \bibnamefont {Lukin}}, \bibinfo {author}
  {\bibfnamefont {H.}~\bibnamefont {Park}}, \ and\ \bibinfo {author}
  {\bibfnamefont {R.~L.}\ \bibnamefont {Walsworth}},\ }\href@noop {} {\bibfield
   {journal} {\bibinfo  {journal} {arXiv preprint arXiv:1705.08887}\ }
  (\bibinfo {year} {2017})}\BibitemShut {NoStop}%
\bibitem [{\citenamefont {Boss}\ \emph {et~al.}(2017)\citenamefont {Boss},
  \citenamefont {Cujia}, \citenamefont {Zopes},\ and\ \citenamefont
  {Degen}}]{boss2017quantum}%
  \BibitemOpen
  \bibfield  {author} {\bibinfo {author} {\bibfnamefont {J.}~\bibnamefont
  {Boss}}, \bibinfo {author} {\bibfnamefont {K.}~\bibnamefont {Cujia}},
  \bibinfo {author} {\bibfnamefont {J.}~\bibnamefont {Zopes}}, \ and\ \bibinfo
  {author} {\bibfnamefont {C.}~\bibnamefont {Degen}},\ }\href@noop {}
  {\bibfield  {journal} {\bibinfo  {journal} {Science}\ }\textbf {\bibinfo
  {volume} {356}},\ \bibinfo {pages} {837} (\bibinfo {year}
  {2017})}\BibitemShut {NoStop}%
\bibitem [{\citenamefont {Laraoui}\ \emph {et~al.}(2013)\citenamefont
  {Laraoui}, \citenamefont {Dolde}, \citenamefont {Burk}, \citenamefont
  {Reinhard}, \citenamefont {Wrachtrup},\ and\ \citenamefont
  {Meriles}}]{laraoui2013high}%
  \BibitemOpen
  \bibfield  {author} {\bibinfo {author} {\bibfnamefont {A.}~\bibnamefont
  {Laraoui}}, \bibinfo {author} {\bibfnamefont {F.}~\bibnamefont {Dolde}},
  \bibinfo {author} {\bibfnamefont {C.}~\bibnamefont {Burk}}, \bibinfo {author}
  {\bibfnamefont {F.}~\bibnamefont {Reinhard}}, \bibinfo {author}
  {\bibfnamefont {J.}~\bibnamefont {Wrachtrup}}, \ and\ \bibinfo {author}
  {\bibfnamefont {C.~A.}\ \bibnamefont {Meriles}},\ }\href@noop {} {\bibfield
  {journal} {\bibinfo  {journal} {Nature communications}\ }\textbf {\bibinfo
  {volume} {4}},\ \bibinfo {pages} {1651} (\bibinfo {year} {2013})}\BibitemShut
  {NoStop}%
\bibitem [{\citenamefont {Zaiser}\ \emph {et~al.}(2016)\citenamefont {Zaiser},
  \citenamefont {Rendler}, \citenamefont {Jakobi}, \citenamefont {Wolf},
  \citenamefont {Lee}, \citenamefont {Wagner}, \citenamefont {Bergholm},
  \citenamefont {Schulte-Herbr{\"u}ggen}, \citenamefont {Neumann},\ and\
  \citenamefont {Wrachtrup}}]{zaiser2016enhancing}%
  \BibitemOpen
  \bibfield  {author} {\bibinfo {author} {\bibfnamefont {S.}~\bibnamefont
  {Zaiser}}, \bibinfo {author} {\bibfnamefont {T.}~\bibnamefont {Rendler}},
  \bibinfo {author} {\bibfnamefont {I.}~\bibnamefont {Jakobi}}, \bibinfo
  {author} {\bibfnamefont {T.}~\bibnamefont {Wolf}}, \bibinfo {author}
  {\bibfnamefont {S.-Y.}\ \bibnamefont {Lee}}, \bibinfo {author} {\bibfnamefont
  {S.}~\bibnamefont {Wagner}}, \bibinfo {author} {\bibfnamefont
  {V.}~\bibnamefont {Bergholm}}, \bibinfo {author} {\bibfnamefont
  {T.}~\bibnamefont {Schulte-Herbr{\"u}ggen}}, \bibinfo {author} {\bibfnamefont
  {P.}~\bibnamefont {Neumann}}, \ and\ \bibinfo {author} {\bibfnamefont
  {J.}~\bibnamefont {Wrachtrup}},\ }\href@noop {} {\bibfield  {journal}
  {\bibinfo  {journal} {Nature Communications}\ }\textbf {\bibinfo {volume}
  {7}} (\bibinfo {year} {2016})}\BibitemShut {NoStop}%
\bibitem [{\citenamefont {Staudacher}\ \emph {et~al.}(2015)\citenamefont
  {Staudacher}, \citenamefont {Raatz}, \citenamefont {Pezzagna}, \citenamefont
  {Meijer}, \citenamefont {Reinhard}, \citenamefont {Meriles},\ and\
  \citenamefont {Wrachtrup}}]{staudacher2015probing}%
  \BibitemOpen
  \bibfield  {author} {\bibinfo {author} {\bibfnamefont {T.}~\bibnamefont
  {Staudacher}}, \bibinfo {author} {\bibfnamefont {N.}~\bibnamefont {Raatz}},
  \bibinfo {author} {\bibfnamefont {S.}~\bibnamefont {Pezzagna}}, \bibinfo
  {author} {\bibfnamefont {J.}~\bibnamefont {Meijer}}, \bibinfo {author}
  {\bibfnamefont {F.}~\bibnamefont {Reinhard}}, \bibinfo {author}
  {\bibfnamefont {C.}~\bibnamefont {Meriles}}, \ and\ \bibinfo {author}
  {\bibfnamefont {J.}~\bibnamefont {Wrachtrup}},\ }\href@noop {} {\bibfield
  {journal} {\bibinfo  {journal} {Nature communications}\ }\textbf {\bibinfo
  {volume} {6}} (\bibinfo {year} {2015})}\BibitemShut {NoStop}%
\bibitem [{\citenamefont {Ajoy}\ \emph {et~al.}(2015)\citenamefont {Ajoy},
  \citenamefont {Bissbort}, \citenamefont {Lukin}, \citenamefont {Walsworth},\
  and\ \citenamefont {Cappellaro}}]{ajoy2015atomic}%
  \BibitemOpen
  \bibfield  {author} {\bibinfo {author} {\bibfnamefont {A.}~\bibnamefont
  {Ajoy}}, \bibinfo {author} {\bibfnamefont {U.}~\bibnamefont {Bissbort}},
  \bibinfo {author} {\bibfnamefont {M.~D.}\ \bibnamefont {Lukin}}, \bibinfo
  {author} {\bibfnamefont {R.~L.}\ \bibnamefont {Walsworth}}, \ and\ \bibinfo
  {author} {\bibfnamefont {P.}~\bibnamefont {Cappellaro}},\ }\href@noop {}
  {\bibfield  {journal} {\bibinfo  {journal} {Physical Review X}\ }\textbf
  {\bibinfo {volume} {5}},\ \bibinfo {pages} {011001} (\bibinfo {year}
  {2015})}\BibitemShut {NoStop}%
\bibitem [{\citenamefont {Rosskopf}\ \emph {et~al.}(2016)\citenamefont
  {Rosskopf}, \citenamefont {Zopes}, \citenamefont {Boss},\ and\ \citenamefont
  {Degen}}]{rosskopf2016quantum}%
  \BibitemOpen
  \bibfield  {author} {\bibinfo {author} {\bibfnamefont {T.}~\bibnamefont
  {Rosskopf}}, \bibinfo {author} {\bibfnamefont {J.}~\bibnamefont {Zopes}},
  \bibinfo {author} {\bibfnamefont {J.}~\bibnamefont {Boss}}, \ and\ \bibinfo
  {author} {\bibfnamefont {C.}~\bibnamefont {Degen}},\ }\href@noop {}
  {\bibfield  {journal} {\bibinfo  {journal} {arXiv preprint arXiv:1610.03253}\
  } (\bibinfo {year} {2016})}\BibitemShut {NoStop}%
\bibitem [{\citenamefont {Laraoui}\ \emph {et~al.}(2011)\citenamefont
  {Laraoui}, \citenamefont {Hodges}, \citenamefont {Ryan},\ and\ \citenamefont
  {Meriles}}]{laraoui2011diamond}%
  \BibitemOpen
  \bibfield  {author} {\bibinfo {author} {\bibfnamefont {A.}~\bibnamefont
  {Laraoui}}, \bibinfo {author} {\bibfnamefont {J.~S.}\ \bibnamefont {Hodges}},
  \bibinfo {author} {\bibfnamefont {C.~A.}\ \bibnamefont {Ryan}}, \ and\
  \bibinfo {author} {\bibfnamefont {C.~A.}\ \bibnamefont {Meriles}},\
  }\href@noop {} {\bibfield  {journal} {\bibinfo  {journal} {Physical Review
  B}\ }\textbf {\bibinfo {volume} {84}},\ \bibinfo {pages} {104301} (\bibinfo
  {year} {2011})}\BibitemShut {NoStop}%
\bibitem [{\citenamefont {Pfender}\ \emph {et~al.}(2016)\citenamefont
  {Pfender}, \citenamefont {Aslam}, \citenamefont {Sumiya}, \citenamefont
  {Onoda}, \citenamefont {Neumann}, \citenamefont {Isoya}, \citenamefont
  {Meriles},\ and\ \citenamefont {Wrachtrup}}]{pfender2016nonvolatile}%
  \BibitemOpen
  \bibfield  {author} {\bibinfo {author} {\bibfnamefont {M.}~\bibnamefont
  {Pfender}}, \bibinfo {author} {\bibfnamefont {N.}~\bibnamefont {Aslam}},
  \bibinfo {author} {\bibfnamefont {H.}~\bibnamefont {Sumiya}}, \bibinfo
  {author} {\bibfnamefont {S.}~\bibnamefont {Onoda}}, \bibinfo {author}
  {\bibfnamefont {P.}~\bibnamefont {Neumann}}, \bibinfo {author} {\bibfnamefont
  {J.}~\bibnamefont {Isoya}}, \bibinfo {author} {\bibfnamefont
  {C.}~\bibnamefont {Meriles}}, \ and\ \bibinfo {author} {\bibfnamefont
  {J.}~\bibnamefont {Wrachtrup}},\ }\href@noop {} {\bibfield  {journal}
  {\bibinfo  {journal} {arXiv preprint arXiv:1610.05675}\ } (\bibinfo {year}
  {2016})}\BibitemShut {NoStop}%
\bibitem [{\citenamefont {Gefen}\ \emph {et~al.}(2018)\citenamefont {Gefen},
  \citenamefont {Khodas}, \citenamefont {McGuinness}, \citenamefont {Jelezko},\
  and\ \citenamefont {Retzker}}]{gefen2018quantum}%
  \BibitemOpen
  \bibfield  {author} {\bibinfo {author} {\bibfnamefont {T.}~\bibnamefont
  {Gefen}}, \bibinfo {author} {\bibfnamefont {M.}~\bibnamefont {Khodas}},
  \bibinfo {author} {\bibfnamefont {L.~P.}\ \bibnamefont {McGuinness}},
  \bibinfo {author} {\bibfnamefont {F.}~\bibnamefont {Jelezko}}, \ and\
  \bibinfo {author} {\bibfnamefont {A.}~\bibnamefont {Retzker}},\ }\href@noop
  {} {\bibfield  {journal} {\bibinfo  {journal} {Physical Review A}\ }\textbf
  {\bibinfo {volume} {98}},\ \bibinfo {pages} {013844} (\bibinfo {year}
  {2018})}\BibitemShut {NoStop}%
\bibitem [{\citenamefont {Casanova}\ \emph {et~al.}(2019)\citenamefont
  {Casanova}, \citenamefont {Torrontegui}, \citenamefont {Plenio},
  \citenamefont {Garc\'{\i}a-Ripoll},\ and\ \citenamefont
  {Solano}}]{casanova2019}%
  \BibitemOpen
  \bibfield  {author} {\bibinfo {author} {\bibfnamefont {J.}~\bibnamefont
  {Casanova}}, \bibinfo {author} {\bibfnamefont {E.}~\bibnamefont
  {Torrontegui}}, \bibinfo {author} {\bibfnamefont {M.~B.}\ \bibnamefont
  {Plenio}}, \bibinfo {author} {\bibfnamefont {J.~J.}\ \bibnamefont
  {Garc\'{\i}a-Ripoll}}, \ and\ \bibinfo {author} {\bibfnamefont
  {E.}~\bibnamefont {Solano}},\ }\href@noop {} {\bibfield  {journal} {\bibinfo
  {journal} {Phys. Rev. Lett.}\ }\textbf {\bibinfo {volume} {122}},\ \bibinfo
  {pages} {010407} (\bibinfo {year} {2019})}\BibitemShut {NoStop}%
\bibitem [{\citenamefont {Cohen-Tannoudji}\ \emph {et~al.}(1986)\citenamefont
  {Cohen-Tannoudji}, \citenamefont {Diu},\ and\ \citenamefont
  {Laloe}}]{cohen1986quantum}%
  \BibitemOpen
  \bibfield  {author} {\bibinfo {author} {\bibfnamefont {C.}~\bibnamefont
  {Cohen-Tannoudji}}, \bibinfo {author} {\bibfnamefont {B.}~\bibnamefont
  {Diu}}, \ and\ \bibinfo {author} {\bibfnamefont {F.}~\bibnamefont {Laloe}},\
  }\href@noop {} {\bibfield  {journal} {\bibinfo  {journal} {Quantum Mechanics,
  Volume 2, by Claude Cohen-Tannoudji, Bernard Diu, Frank Laloe, pp. 626. ISBN
  0-471-16435-6. Wiley-VCH, June 1986.}\ ,\ \bibinfo {pages} {626}} (\bibinfo
  {year} {1986})}\BibitemShut {NoStop}%
\bibitem [{\citenamefont {Gillespie}(1996)}]{gillespie1996exact}%
  \BibitemOpen
  \bibfield  {author} {\bibinfo {author} {\bibfnamefont {D.~T.}\ \bibnamefont
  {Gillespie}},\ }\href@noop {} {\bibfield  {journal} {\bibinfo  {journal}
  {Physical review E}\ }\textbf {\bibinfo {volume} {54}},\ \bibinfo {pages}
  {2084} (\bibinfo {year} {1996})}\BibitemShut {NoStop}%
\bibitem [{\citenamefont {Aharon}\ \emph {et~al.}(2016)\citenamefont {Aharon},
  \citenamefont {Cohen}, \citenamefont {Jelezko},\ and\ \citenamefont
  {Retzker}}]{aharon2016fully}%
  \BibitemOpen
  \bibfield  {author} {\bibinfo {author} {\bibfnamefont {N.}~\bibnamefont
  {Aharon}}, \bibinfo {author} {\bibfnamefont {I.}~\bibnamefont {Cohen}},
  \bibinfo {author} {\bibfnamefont {F.}~\bibnamefont {Jelezko}}, \ and\
  \bibinfo {author} {\bibfnamefont {A.}~\bibnamefont {Retzker}},\ }\href@noop
  {} {\bibfield  {journal} {\bibinfo  {journal} {New J. Phys.}\ }\textbf
  {\bibinfo {volume} {18}},\ \bibinfo {pages} {123012} (\bibinfo {year}
  {2016})}\BibitemShut {NoStop}%
\end{thebibliography}%
\bibliographystyle{apsrev4-1}

\newpage

\section{Supplementary Material}

\section{The Model}
We consider an NV center electronic spin that is interacting with a single or several nuclei via the dipole - dipole interaction.
As we are interested in the regime in which the energy gap due to the Zeeman splitting of the NV is orders of magnitude larger than the energy gap of the nucleus, 
only the $T_{+1} + T_{-1} = \frac{3}{2} \sin \theta \cos \theta  \sigma_z \left(I_x \cos \phi + I_y \sin \phi \right),$ (see, for example, \cite{cohen1986quantum}, Complement $B_{XI}$) term of the dipole - dipole interaction is significant, where $\theta, \phi$ are the polar angles representing the vector joining the NV and the nucleus.
All other terms of the dipole-dipole interaction are fast rotating and thus can be neglected to leading order. In most cases it is the above term that is used for polarization and for sensing, in particular, in the high-field NMR experiments. Because the energy gaps of the  ground state sub-levels of the NV are much larger than the RF, only two levels are addressed by the microwave driving fields and thus, the NV could be approximated as a two-level system. 
\section{The Hartmann-Hahn condition}
\label{sec_HH}
Under an on-resonance drive of the NV, the Hamiltonian of the NV center spin and the nuclear spin is given by
\begin{equation}
H = \frac{\omega_0}{2} \sigma_z + \frac{\omega_l}{2} I_z + g \sigma_z I_x + \Omega_1 \sigma_x \cos \left(\omega_0 t \right),
\label{HH1}
\end{equation}
where $\omega_0$ corresponds to the energy gap of the NV center spin, $\omega_l$ is the Larmor frequency of the nuclear spin,  $\sigma_z$ and $I_z$ are the Pauli operators in the direction of the static magnetic field of the NV center and the nucleus respectively, $g$ is the NV - nucleus coupling strength, which depends on the distance between the two, and where we have simplified the $T_1+T_{-1}$ term to $g \sigma_z I_x$, and $\Omega_1$ is the Rabi frequency    of the on-resonance driving field of the NV center.  By moving to the interaction picture (IP) with respect to the first term, $H_0 =  \frac{\omega_0}{2} \sigma_z $, and making the rotating-wave-approximation (RWA) assuming that $\omega_0 \gg \Omega_1$, we obtain
\begin{equation}
H_I =  \frac{\Omega_1}{2} \sigma_x + \frac{\omega_l}{2} I_z + g \sigma_z I_x . 
\end{equation}
In the basis of the dressed NV center states ($x \rightarrow z$, $z \rightarrow -x$, and $y \rightarrow y$) we have that
\begin{equation}
H_I =  \frac{\Omega_1}{2} \sigma_z + \frac{\omega_l}{2} I_z - g \sigma_x I_x. 
\end{equation} 
Moving now to the second IP with respect to $H_0 =  \frac{\Omega_1}{2} \sigma_z + \frac{\omega_l}{2} I_z$, we arrive at
\begin{equation}
H_{II} \approx  - g\left(\sigma_+  I_- e^{i\left(\Omega_1-\omega_l\right) t} + \sigma_- I_+ e^{-i\left(\Omega_1-\omega_l\right) t}\right), 
\label{HH}
\end{equation} 
where fast rotating terms have been neglected. It is clear from  eq. \ref{HH} that for sensing and control of the nuclear spin by the NV it is necessary to fulfill  the resonance condition, $\Omega_1 = \omega_l$, which is the Hartmann-Hahn condition. 

\begin{figure}[t]
\begin{center}
\includegraphics[width=0.48\textwidth]{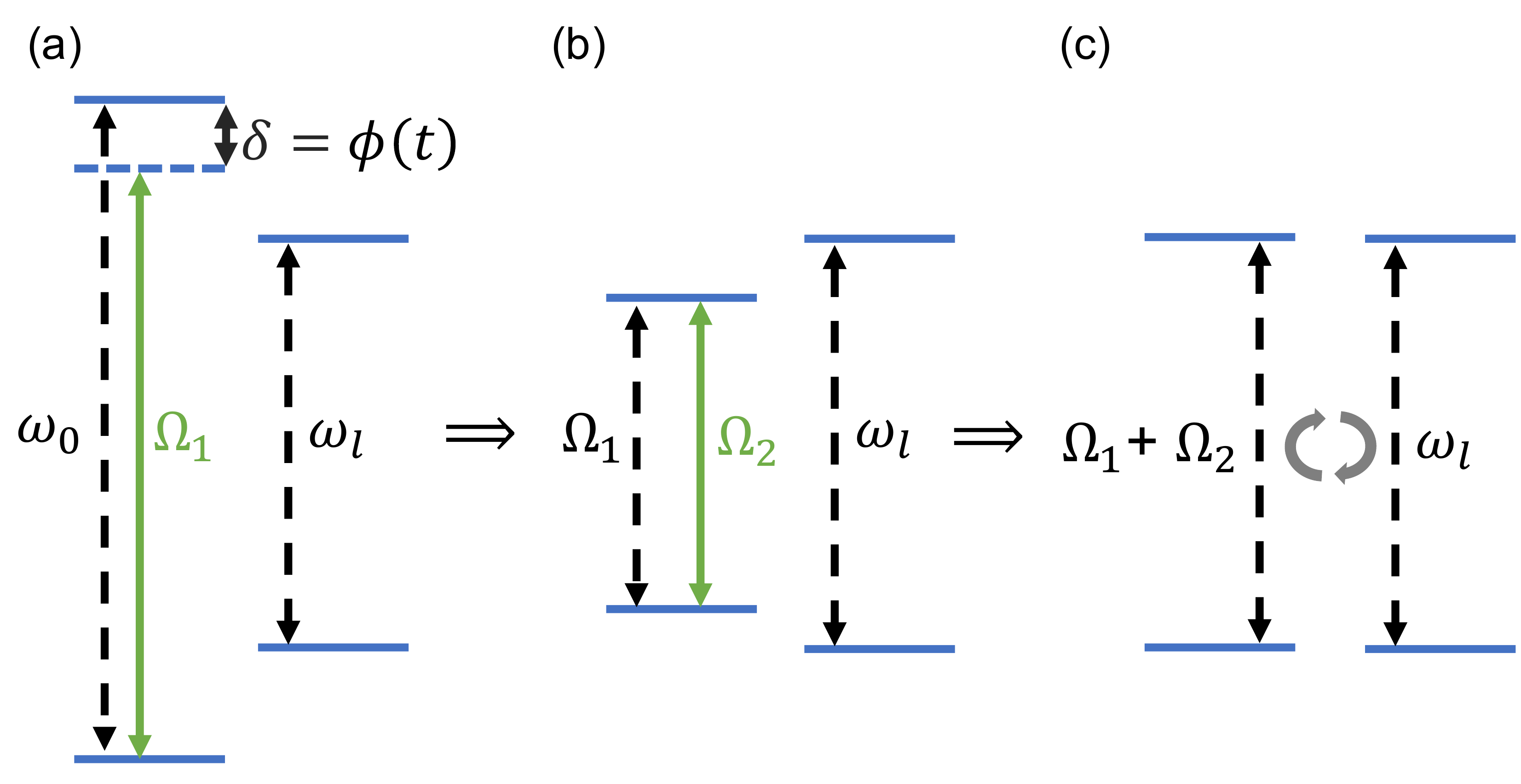}
\end{center}
\caption{Small frequency mismatch scheme. (a) The  electron spin is driven with a bounded RF $\Omega_1$ that is smaller than the nuclear LF $\omega_l$ ($\Omega_1<\omega_l$), but with a phase modulation $\phi\left(t\right)$ (see text). (b) This results in electronic dressed states with an energy gap of $\Omega_1$ that are driven on-resonance by a second drive with a RF of $\Omega_2$. The second drive originates only from the phase modulation which does not require additional power beyond the power of $\propto \Omega_1^2$ that is required for the bounded RF of $\Omega_1$. (c) The second drive $\Omega_2$ results in double dressed states of the electron that match the resonance condition with $\Omega_1 + \Omega_2 = \omega_l$. Doted lines indicate resonance frequencies and solid lines indicate driving fields.}
\label{smallM}
\end{figure}

\section{Phase modulation - the basic scheme}
We consider the following Hamiltonian of the NV center and the nucleus, 
\begin{eqnarray}
H &=& \frac{\omega_0}{2} \sigma_z + \delta B(t) \sigma_z+ \frac{\omega_l}{2} I_z + g \sigma_z I_x  \nonumber \\
 &+& \left( \Omega_1 + \delta \Omega_1(t)  \right) \sigma_x \cos \left( \omega_0 t  + 2 \frac{\Omega_2}{\Omega_1} \sin(\Omega_1 t)  \right),
 \label{phasedep}
\end{eqnarray}
where $ \delta B(t) $ is the noise in the magnetic field,  $\Omega_1$  is the RF of the driving field, which defines the phase 
modulation according to $\phi\left(t\right)=2 \frac{\Omega_2}{\Omega_1} \sin\left(\Omega_1 t\right) $, and $ \delta \Omega_1(t) $ is the amplitude noise in the drive amplitude $\Omega_1$.

In order to see how the Hamiltonian of Eq. \ref{phasedep} results in the resonance condition $\Omega_1 + \Omega_2 = \omega_l$, we start by moving to the first IP in which the drive is time indepandant, i.e., with respect to $H_0 = \frac{\omega_0 + 2 \Omega_2 \cos(\Omega_1 t)}{2}\sigma_z$.
This results in
\begin{eqnarray}
H_I &=& \frac{\left( \Omega_1 + \delta \Omega_1(t)  \right)}{2} \sigma_x + \delta B(t) \sigma_z -  \Omega_2 \cos \left( \Omega_1 t \right)  \sigma_z \nonumber\\
&+& \frac{\omega_l}{2} I_z + g \sigma_z I_x,
\label{eqIP1}
\end{eqnarray}
which is similar to a concatenated double-drive Hamiltonian, this time, however, with a very stable second drive, $\Omega_2$.  Because the magnetic noise is perpendicular to the basis of the dressed states robustness to the magnetic noise (in first order) is achieved. From here on, we neglect the magnetic noise whose leading (second order) contribution  is $\sim \frac{ \delta B(t)^2}{\Omega_1}$.

We continue by rotating to the basis of the dressed states (as in section \ref{sec_HH}) such that
\begin{eqnarray}
H_I &=& \frac{\left( \Omega_1 + \delta \Omega_1(t)  \right)}{2} \sigma_z +\Omega_2 \cos \left( \Omega_1 t \right)  \sigma_x \nonumber\\
&+& \frac{\omega_l}{2} I_z - g \sigma_x I_x.
\end{eqnarray}
It is now clear that the phase modulation results in a second drive that drives the dressed states on-resonance with a RF of $\Omega_2$. The double-dressed states are obtained by moving to the second IP with respect to $H_0=\frac{ \Omega_1}{2} \sigma_z$, 
\begin{eqnarray}
H_{II} &=& \frac {\Omega_2}{2}\sigma_x+ \frac{\delta \Omega_1(t)}{2} \sigma_z \nonumber\\
&+& \frac{\omega_l}{2} I_z - g\left(  \sigma_+ e^{i \Omega_1 t} + \sigma_- e^{-i \Omega_1 t} \right) I_x.
\label{IP2}
\end{eqnarray}
Because the amplitude noise $ \delta \Omega_1(t) $ is perpendicular to the basis of the double-dressed states, robustness to the amplitude noise (in first order) is achieved. From here on, we neglect the amplitude noise whose leading (second order) contribution  is $\sim \frac{\delta \Omega_1(t) ^2}{\Omega_2}$.
Moving to the basis of the double-dressed states we get
\begin{eqnarray}
\label{eq7}
H_{II} &=& \frac {\Omega_2}{2}\sigma_z+ \frac{\omega_l}{2} I_z - g\left( \cos \left(\Omega_1 t\right) \sigma_z + \sin \left(\Omega_1 t\right) \sigma_y\right) I_x \nonumber\\ 
&\approx& \frac {\Omega_2}{2}\sigma_z+ \frac{\omega_l}{2} I_z \\
&-& \frac{g}{2} \left(  \sigma_{+} \left(e^{i \Omega_1 t}- e^{-i \Omega_1 t}\right)+ \sigma_{-} \left(e^{-i \Omega_1 t} - e^{i \Omega_1 t}\right) \right) I_x \nonumber,
\end{eqnarray}
where we have omitted the fast rotating terms $g \cos \left(\Omega_1 t\right) \sigma_z $ in the approximation.
From this expression it is seen that a resonance condition appears when $\Omega_1 +\Omega_2  = \omega_l$
(or when  $\Omega_1 -\Omega_2  = \omega_l$).
Even though the power of the driving field is $\propto\Omega_1^2$ and is independent of $\Omega_2,$ Larmor frequencies which are higher than what is available by the peak power in a common HH scheme are reachable  (Fig. \ref{smallM}).

\section{Correction of the Bloch-Siegert shift}
In this section we give a detailed derivation of the correction of the Bloch-Siegert shift. The correction can be understood as  follows. Without the correction, we first consider the dressed states due to the rotating-terms of the drive ($\frac{\Omega_2}{2}\sigma_x$) and then consider the effect of the off-resonance counter-rotating terms of the drive ($\frac{\Omega_2}{2}\left(\sigma_+ e^{i \Omega_1 t} + \sigma_- e^{-i \Omega_1 t}\right)$) on the dressed states (the eigenstates  of $\frac{\Omega_2}{2}\sigma_x$). This results in an energy shift of the dressed states, and  (a time-dependent) amplitude-mixing between the dressed states, which decreases the coherence time.

To correct this effect, we first consider the effect of the counter-rotating terms on the bare states, and then fix the frequency of the drive accordingly such that the rotating-terms will be on-resonance with the modified bare states. Consider the driving Hamiltonian
\begin{equation}
H_d= \frac{\Omega_1}{2} \sigma_x -\Omega_2 \cos \left( \omega_2 t \right)  \sigma_z.
\end{equation} 
Instead of moving to the IP of the rotating frame we first move to the IP of the counter-rotating frame with respect to $H_0 = -\frac{\omega_2}{2} \sigma_x$ and obtain
\begin{equation}
H_I= \frac{\Omega_1+\omega_2}{2} \sigma_x -\frac{\Omega_2}{2}\sigma_z   -\frac{\Omega_2}{2}\left(\sigma_+ e^{-2i\omega_2 t}+\sigma_- e^{+2i\omega_2} \right).
\end{equation} 
We continue by moving to the diagonal basis of the time-independent part of $H_I$, 
\begin{equation}
H_I\approx \frac{1}{2} \sqrt{(\Omega_1+\omega_2)^2+\Omega_2^2} \sigma_z   -\frac{\tilde{\Omega}_2}{2}\left( \sigma_+ e^{-2i\omega_2 t}+\sigma_- e^{+2i\omega_2} \right),
\end{equation} 
where
\begin{equation}
\tilde{\Omega}_2 = \frac{\Omega_2}{2}\left(1+\frac{\Omega_1+\omega_2}{\sqrt{\Omega_2^2+\left(\Omega_1+\omega_2\right)^2}}\right).
\label{eq_tilde_O2}
\end{equation}
By setting $2\omega_2= \sqrt{(\Omega_1+\omega_2)^2+\Omega_2^2}$ we have that the rotating terms are on-resonance with the energy gap of the modified bare states. The on-resonance condition is therefore given by 
\begin{equation}
\omega_2=\frac{1}{3}\left(\Omega_1+ \sqrt{4 \Omega_1^2+3 \Omega_2^2}\right).
\end{equation}
In this case the amplitude-mixing between the dressed states is greatly diminished and hence, the coherence time of the NV center may be significantly prolonged compared to the scenario without the correction. Hence, we modify the frequency $\Omega_1$ in the phase modulation $\phi\left(t\right)=2 \frac{\Omega_2}{\Omega_1} \sin\left(\Omega_1 t\right) $  in Eq. \ref{phasedep} to
\begin{equation}
\tilde{\Omega}_1 = \frac{1}{3}\left( \Omega_1+ \sqrt{4 \Omega_1^2+3 \Omega_2^2}\right).
\label{eq_tilde_O1}
\end{equation}
Eq. \ref{eq_tilde_O2} and Eq. \ref{eq_tilde_O1} imply that the resonance frequency, which is given by $\Omega_1 + \Omega_2 = \omega_l$, is modified to 
\begin{equation}
\tilde{\Omega}_1 +\tilde{\Omega}_2  = \omega_l.
\end{equation}

\section{Numerical analysis}
\subsection{Strong coupling regime}

\begin{figure}[t!]
\begin{center}
\includegraphics[width=0.48\textwidth]{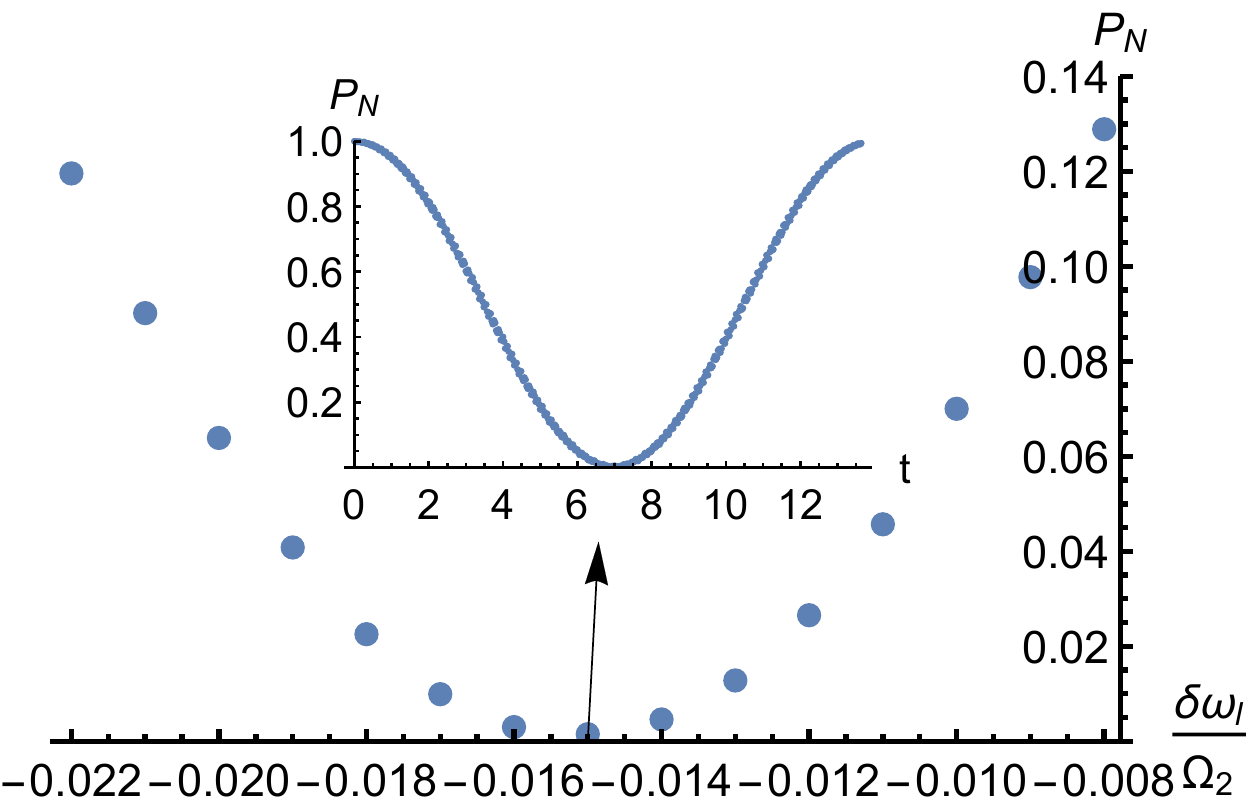}
\end{center}
\caption{The polarization, which is defined as the population of the nuclei in the initial $\ket \uparrow_z$ state (see text), as a function of the resonance frequency shift $\delta \omega_l$ in units of $\Omega_2$. Here we consider polarization with the correction of the Bloch-Siegert shift with $\Omega_2 = 1.2 \Omega_1$. The optimal polarization is obtained for $\frac{\delta \omega_l}{\Omega_2} = -0.015$ and is equal to $P_N = 0.0015$. (Inset) The polarization as function of time for the optimal value of  $\delta \omega_l$. The time is in units of $\mu$s.}
\label{Polar1}
\end{figure}

In the strong coupling regime we consider the scenario in which the polarization time $t=2\pi/g$ is short enough such that the effect of noise on the polarization is minor, that is, the polarization time is much shorter than the decoherence time of the NV center. Hence, we neglect decoherence effects. 

\begin{figure}[t!]
\begin{center}
\includegraphics[width=0.48\textwidth]{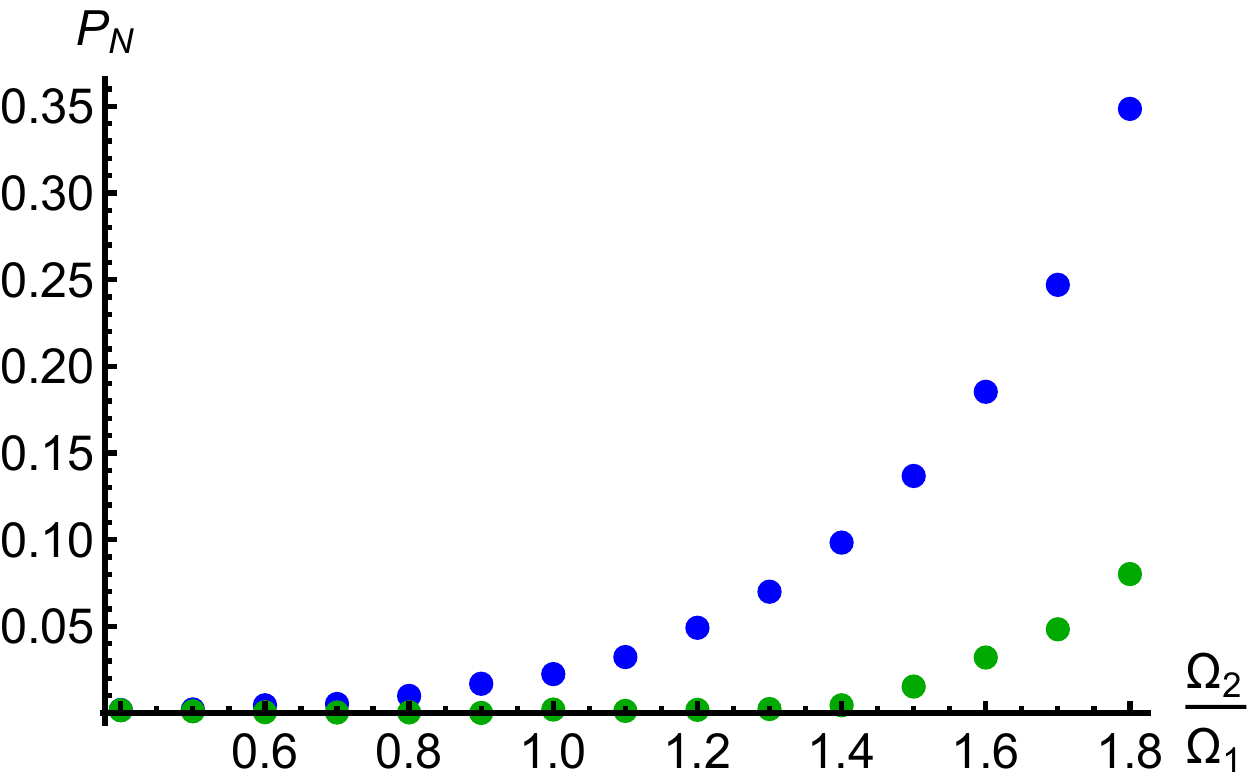}
\end{center}
\caption{The polarization as a function of $\Omega_2$ in units of $\Omega_1$ in the strong coupling regime.   First method without correction of the Bloch-Siegert shift (blue) -  the polarization rate begins to sharply decrease at $\Omega_2 \approx \Omega_1$. Second method with correction of the Bloch-Siegert shift (green) -  the correction enables to maintain good polarization rates up to $\Omega_2 \approx 1.8\Omega_1$.}
\label{Polar2}
\end{figure}

Because we consider the regime of high magnetic fields, we have that $\omega_0 \gg \Omega_1$ and the RWA is valid with respect to the first drive $\Omega_1$. Hence, in the numerical analysis we simulated the Hamiltonian in the first IP, which is given by Eq. \ref{eqIP1}. Since here we neglect decoherence effects we omitted the terms of the magnetic and drive noise. The simulations were performed with $\Omega_1 = 2 \pi \times 3.3$ MHz and $g = 0.04 \Omega_1$, where the value of $\Omega_2$ was varied. For each value of $\Omega_2$ we scanned the resonance frequency around the ideal value of $\omega_l = \Omega_1 + \Omega_2$ or $\omega_l =\tilde{\Omega}_1 +\tilde{\Omega}_2$ (without and with the correction of the Bloch-Siegert shift respectively) and found the additional shift in the resonance frequency $\delta \omega_l$, which results from the effect of the fast-rotating terms (Fig. \ref{Polar1}). At the resonance frequency,  $\omega_l = \Omega_1 + \Omega_2 + \delta \omega_l$ (or $\omega_l = \tilde{\Omega}_1 + \tilde{\Omega}_2 + \delta \omega_l$ with the correction), the maximal polarization is obtained. We define the nuclear spin polarization, $P_N$, as the probability of the nuclear spin to be in its initial state $\ket{\uparrow_z}$. Specifically, we initialize the NV-Nucleus state to    $\ket{\psi_i}=\ket{\downarrow_z}_{NV}\ket{\uparrow_z}_{N}=\ket{\downarrow_z \uparrow_z}$ and calculate the polarization according to $P_N =|\braket{\uparrow_z \uparrow_z}{\psi}|^2 + |\braket{\downarrow_z \uparrow_z}{\psi}|^2$, where $\ket{\psi}$ is the joint NV-Nucleus state at the optimal polarization time. Hence, $P_N=0$ corresponds to optimal polarization and $P_N=1$ corresponds to no polarization at all. 
Note that since the $\hat{z}$ basis is the basis of the double-dressed states and also the measurement axis, the Rabi frequencies $\Omega_1$ and $\Omega_2$ are invisible to the population measurement of $P_N$.   
In Fig. (\ref{Polar2}) we show the nuclei polarization as function of the ratio $\frac{\Omega_2}{\Omega_1}$ both with (green) and without (blue) the correction of the Bloch-Siegert shift. It is clear that in the strong coupling regime the correction results in better polarization rates, especially when $\Omega_2 \gtrsim \Omega_1$.

\subsection{Weak coupling regime}
\begin{figure}[t!]
\begin{center}
\includegraphics[width=0.48\textwidth]{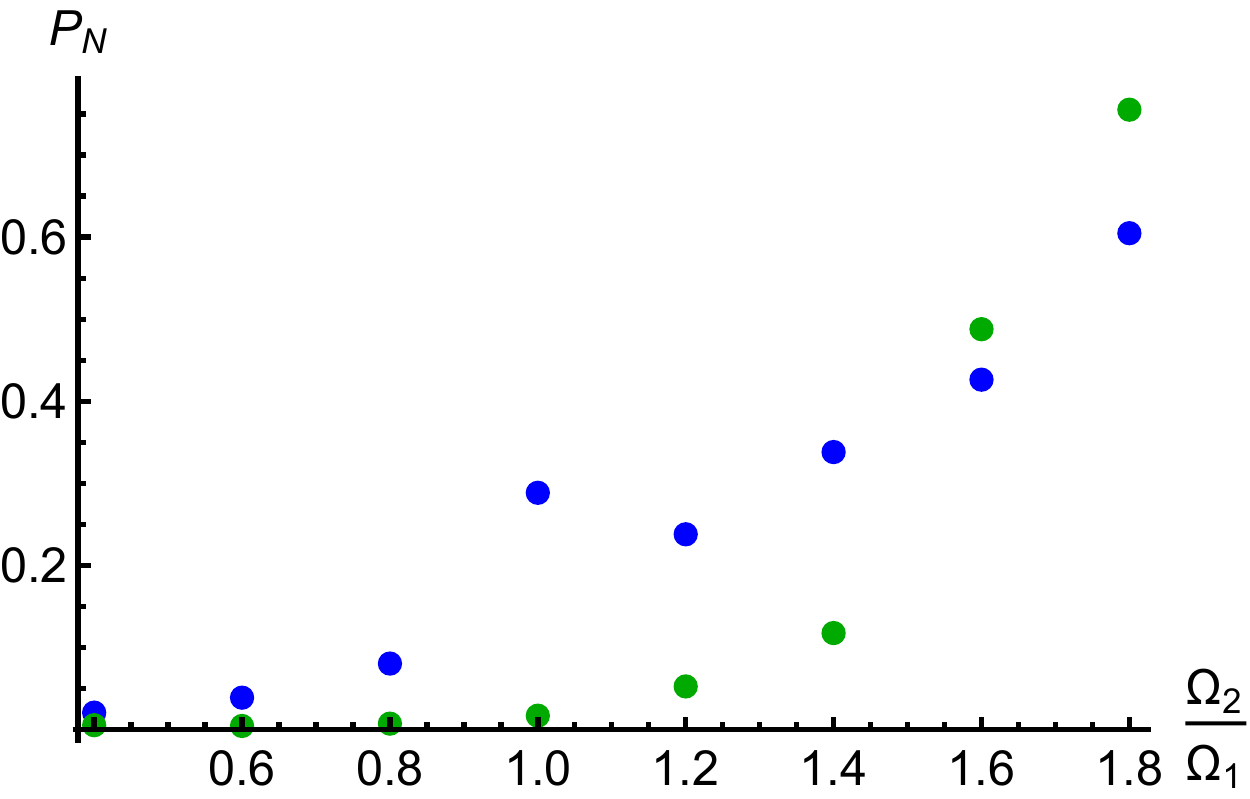}
\end{center}
\caption{The polarization as a function of $\Omega_2$ in units of $\Omega_1$ in the weak coupling regime.   First method without correction of the Bloch-Siegert shift (blue). Second method with correction of the Bloch-Siegert shift (green). The analysis takes noise into account, which is crucial for a weak coupling ($g$). The polarization is effective upto $\Omega_2 \approx 1.4 \Omega_1$.}
\label{Polar3}
\end{figure}

In the weak coupling regime the polarization time $t=2\pi/g$ is long enough such that decoherence effects must be taken into account. Hence, the terms of the magnetic noise and the driving amplitude noise in  Eq. \ref{eqIP1} are not omitted. In Fig. (\ref{Polar3}) we show the nuclei polarization as function of the ratio $\frac{\Omega_2}{\Omega_1}$ both with (green) and without (blue) the correction of the Bloch-Siegert shift.  The simulations were performed with $\Omega_1 = 2 \pi \times 3.3$ MHz and $g = 0.01 \Omega_1$. While in the strong coupling regime the correction always results in better polarization rates, in the weak coupling regime the advantage of correction is lost at $\Omega_2 \approx 1.5 \Omega_1$. 


The numerical simulations of the polarization rates and coherence times were performed under the assumption that the pure dephasing time of the NV is $T_2^*= 3 \mu$s, which results from a magnetic noise, $B\left(t\right)$,  that is modeled by an Ornstein-Uhlenbeck (OU) process \cite{gillespie1996exact, aharon2016fully} with a zero expectation value, $\left\langle B\left(t\right)\right\rangle =0$,
and a correlation function $\left\langle B\left(t\right)B\left(t'\right)\right\rangle =\frac{c\tau}{2}e^{-\gamma\left|t-t'\right|}$, where $c$ is the diffusion constant and $\tau = \frac{1}{\gamma} = 25 \mu$s is the correlation time of the noise. An OU process was also used to realize driving fluctuations. Here we used a correlation time of
$\tau_{\Omega}=500\:\mu s$, and a relative amplitude error of $\delta_{\Omega}=1\%$.

\section{Large frequency mismatch}
\subsection{Method I - Detuned driving field}
In this section we provide the derivation of the the detuned driving method.
The calculation goes as follows.  By introducing a detuning $\delta$ to the driving filed in Eq. \ref{HH1} we obtain the Hamiltonian 
\begin{equation}
H = \frac{\omega_0}{2} \sigma_z + \frac{\omega_l}{2} I_z + \Omega_1\cos((\omega_0-\delta)t)\sigma_x + g \sigma_z I_x.
\end{equation}
In order to analyze this scenario it is advantageous to move to the IP in which the drive is time independent and the problem can be analyzed  via the resulting dressed states.
Hence, we choose to move to the IP with respect to $ \frac{\omega_0-\delta}{2} \sigma_z$, which results in
\begin{eqnarray}
H_I &=& \frac{\delta}{2} \sigma_z+ \frac{\Omega_1}{2} \sigma_x + \frac{\omega_l}{2} I_z+ g \sigma_z  I_x \\
      &=& \frac{ \sqrt{\delta^2+\Omega_1^2}}{2} \sigma_\theta + \frac{\omega_l}{2} I_z+ g(\sigma_\theta \cos \theta + \sigma_{\theta_{\perp}} \sin \theta)  I_x,\nonumber 
\end{eqnarray}
where $\sigma_{\theta}$ is a Pauli matrix in the direction $\theta = \arctan \left( \frac{\Omega_1}{\delta} \right),$ which is the angle in the $x-z$ plane from the $z$ axis and $\sigma_{\theta_{\perp}}$ is the Pauli matrix in the orthogonal direction.

Close to resonance, when  $\sqrt{\delta^2+\Omega_1^2} = \omega_l,$ we can expect excitation transfer between the electron and the nucleus due to the last term, i.e., $ \sigma_{\theta_{\perp}} \sin (\theta)  I_x$.
Since for large detunings $\sin \theta \approx \frac{\Omega_1}{\delta}$ the effective coupling strength is reduced from $g$ to $\approx \frac{\Omega_1}{\delta} g$, which is depicted in Fig.1 (c) of the main text.

This method, however, suffers from the fact that the decoupling effect due to the resonant drive vanishes and the coherence time of the NV approaches the $T_2^*$ time.  Specifically, when moving to the first IP, the magnetic noise  $\delta B(t) \sigma_z$ is modified to  $\delta B(t)(\sigma_\theta \cos \theta + \sigma_{\theta_{\perp}} \sin \theta) $, which results in a first order contribution as long as has $\cos \theta \neq 0$.  In order to circumvent this issue and to prolong the coherence time we suggest to introduce a second drive by adding an extra term in the Hamiltonian, namely, 
\begin{eqnarray}
H &=& \frac{\omega_0}{2} \sigma_z + \frac{\omega_l}{2} I_z  + g \sigma_z I_x +  \Omega_1\cos((\omega_0-\delta)t)\sigma_x \nonumber\\
   &+& \Omega_2 \cos\left((\omega_0-\delta\right)t + \frac{\pi}{2})\cos\left( \sqrt{\delta^2+\Omega_1^2}\right)\sigma_x.
\end{eqnarray}
Moving to the first IP with respect to  $ \frac{\omega_0-\delta}{2} \sigma_z$  and the to the basis of the dressed states as above, we obtain
\begin{eqnarray}
H_I &=& \frac{ \sqrt{\delta^2+\Omega_1^2}}{2} \sigma_\theta + \frac{\omega_l}{2} I_z+ g(\sigma_\theta \cos \theta + \sigma_{\theta_{\perp}} \sin \theta)  I_x \nonumber\\
&+& \frac{\Omega_2}{2}\cos\left( \sqrt{\delta^2+\Omega_1^2}\right)\sigma_y.
\end{eqnarray}
We continue by moving to the second IP with respect to $H_0=\frac{ \sqrt{\delta^2+\Omega_1^2}}{2} \sigma_\theta$, which leads to
\begin{eqnarray}
H_{II} &\approx & \frac{\Omega_2}{4}\sigma_y+ \frac{\omega_l}{2} I_z \nonumber\\
&+&  g \sin \theta (\sigma_{\theta_{+}} e^{i\sqrt{\delta^2+\Omega_1^2}t }+\sigma_{\theta_{-}} e^{-i\sqrt{\delta^2+\Omega_1^2}t })  I_x,
\end{eqnarray}
where $\sigma_{\theta_{+}}$ and $\sigma_{\theta_{-}}$ are the raising and lowering operators in the basis of $\sigma_{\theta}$ respectively.  Similar to Eq. \ref{IP2}, we see that the resonance condition is fulfilled when $\sqrt{\delta^2 + \Omega_1^2} + \frac{\Omega_2}{2} = \omega_l$. Hence, Larmor frequencies that are much higher than what is available by the power limitation, which here is $\propto \Omega_1^2 + \Omega_2^2$, are reached. 

Because the second  drive $\Omega_2$ is along the $y$  axis, which is perpendicular to the basis of the dressed states that is in the $x-z$ plane, the second drive achieves robustness to (first order) magnetic noise and amplitude noise in $\Omega_1$. Hence, the second drive prolongs the coherence time of the NV, and thus the resolution, while shifting the resonance to $\sqrt{\delta^2 + \Omega_1^2} + \frac{\Omega_2}{2} = \omega_l$. 
The disadvantage of this method is that the second drive $\Omega_2$, which results in a drive along the $y$ direction, cannot be generated by a phase modulation, which results in a drive along the $z$ direction.  and thus it is not robust against amplitude fluctuations of $\Omega_2$. Meaning, this decoupling limit is as good as the coherence time achieved in regular spin locking which is roughly an order of magnitude longer than $T_2^*.$ For some scenarios (weak coupling regime) the coherence time  will have to be further prolonged by coherent control, for example, by adding an extra drive.
  
\subsection{Method II - Amplitude modulation}

With only a single amplitude modulation the method suffers from amplitude fluctuations in $\Omega_0.$
These fluctuations could be eliminated by creating this drive as a second drive from a phase modulation as in the small frequency mismatch method. Specifically, the driving Hamiltonian of the NV center is given by
\begin{equation}
H = \frac{\omega_0}{2} \sigma_z + \Omega_0 \cos \left(\omega_0 t + \varphi\left(t\right)\right), 
\end{equation}
where 
\begin{eqnarray}
\varphi\left(t\right)&=& 2\left(\frac{\Omega_1}{\Omega_0}\sin\left(\Omega_0 t\right)\right.\nonumber\\
&+&\frac{\Omega_2}{\Omega_0^2-\Omega_3^2}\left[\Omega_0\cos\left(\Omega_3 t\right)\sin\left(\Omega_0 t\right)\right.\nonumber\\
&-& \left.\Omega_3\cos\left(\Omega_0 t\right)\sin\left(\Omega_3 t\right)\right]\bigg)
\end{eqnarray}
Moving to the IP with respect to $H_0=\frac{\omega_0+\left(\Omega_1+\Omega_2\cos\left(\Omega_3 t\right)\right)\cos\left(\Omega_0 t\right)}{2} \sigma_z$ we obtain that
\begin{eqnarray}
H_I = \frac{\Omega_0}{2}\sigma_x -\left(\Omega_1+\Omega_2\cos\left(\Omega_3 t\right)\right)\cos\left(\Omega_0 t\right)\sigma_z.
\end{eqnarray}
Hence, robustness to amplitude fluctuation in $\Omega_0$ is achieved. Moreover, due to the utilization of phase modulation, increasing either $\Omega_1$, $\Omega_2$, or $\Omega_3$ is not associated with an increased power consumption.

\section{Power consumption}
In this section we consider the difference in power consumption of the proposed schemes in comparison to the common Hartmann-Hahn method. For a given driving field, the magnitude of the magnetic field is proportional to the Rabi frequency, $B(t) \propto \Omega(t)$, and since the magnitude of the electric field is proportional to the magnitude of the magnetic field we have that the power density $P(t)=\frac{1}{\mu_0}\bold{E}\times\bold{B}$, where $\mu_0$ is the  vacuum permeability, is $P(t) \propto \Omega^2(t)$. Because we are only interested in the ratio between the power consumption of the Hartmann-Hahn method, $P_{HH}(t)$, and the power consumption of a proposed method $m$, $P_{m}(t)$, we have that $\frac{P_{HH}(t)}{P_{m}(t)}=\frac{\Omega_{HH}^2(t)}{\Omega_{m}^2(t)}$, and hence, it is not necessary to calculate the exact power density. 

We consider two figure of merits for the comparison of power consumption. The first is the peak power of a drive (the maximal instantaneous power value), which we denote by $P^{peak}$, and the second is the total cycle power that is required for a complete energy transfer (flip-flop) between the NV spin and the nucleus, which is denoted by $P^{cycle}=\int_{0}^{T}P(t)$, where $T$ is the cycle time. 

The Rabi frequency of an on-resonance Hartmann-Hahn drive is given by $\Omega_{HH}=\Omega \cos(\omega_0 t)$, where, $\Omega = \omega_l$. Denoting the NV-nucleus coupling rate by $g$, the Hartmann-Hahn cycle time is given by $T_{HH}=\frac{2\pi}{g}$. For the Hartmann-Hahn drive we therefore have that 
\begin{equation}
P_{HH}^{peak} \propto \Omega^2,\qquad P_{HH}^{cycle} \propto \frac{1}{2}\Omega^2 T_{HH},
\end{equation}
where for $P_{HH}^{cycle}$ the above expression is approximately correct in the limit of $\omega_0\gg\Omega$.
\subsection{Phase modulation}
The Rabi frequency of a phase modulated driving field is given by $ \Omega_1 \cos \left( \omega_0 t  + 2 \frac{\Omega_2}{\Omega_1} \sin(\Omega_1 t)  \right)$. In the case of phase modulation the effective NV-nucleus coupling rate is reduced by a factor of $2$ (see Eq. (\ref{eq7})) and hence the cycle time is increased by a factor of $2$, $T_{PM} = \frac{4\pi}{g}$, so we have that
\begin{equation}
P_{PM}^{peak} \propto \Omega_1^2,\qquad P_{PM}^{cycle} \propto \frac{1}{2}\Omega_1^2 T_{PM}.
\end{equation}
This results in
\begin{equation}
\frac{P_{HH}^{peak}}{P_{PM}^{peak}} = \frac{\Omega^2}{\Omega_1^2},\qquad \frac{P_{HH}^{cycle}}{P_{PM}^{cycle}} \approx \frac{\Omega^2 T_{HH}}{\Omega_1^2 T_{PM}}.
\end{equation} 
The phase modulation scheme is relevant for the small frequency mismatch regime, where good polarization rates can be achieved for $\Omega_2=\Omega_1$. In this case $\Omega=\omega_l=2 \Omega_1$ and hence
\begin{equation}
\frac{P_{HH}^{peak}}{P_{PM}^{peak}} = \frac{\Omega^2}{\Omega_1^2}=4,\qquad \frac{P_{HH}^{cycle}}{P_{PM}^{cycle}} \approx \frac{\Omega^2 T_{HH}}{\Omega_1^2 T_{PM}}=2.
\end{equation}

\subsection{Detuned driving field}
The Rabi frequency of a detuned driving field is given by $ \Omega_1 \cos \left(\left( \omega_0-\delta\right) t  \right)$. Recall that the effective NV-nucleus coupling rate is reduced by a factor of $\sin \theta \approx \frac{\Omega_1}{\delta}$ from $g$ to $\approx \frac{\Omega_1}{\delta} g$ so the cycle time is increased to $T_{det} = \frac{2 \delta\pi}{\Omega_1 g}$. For a detuned driving field 
\begin{equation}
P_{det}^{peak} \propto \Omega_1^2,\qquad P_{det}^{cycle} \propto \frac{1}{2}\Omega_1^2 T_{det}.
\end{equation}
Assuming, for example, that $\delta = 10 \Omega_1$ so $\omega_l = \sqrt{101}\Omega_1$, which corresponds to the large frequency mismatch, we have that  
\begin{equation}
\frac{P_{HH}^{peak}}{P_{det}^{peak}} = \frac{\Omega^2}{\Omega_1^2}=101,\qquad \frac{P_{HH}^{cycle}}{P_{det}^{cycle}} \approx \frac{\Omega^2 T_{HH}}{\Omega_1^2 T_{det}}=10.1.
\end{equation}

\subsection{Amplitude modulation}
The Rabi frequency of an amplitude modulated driving field is given by $ \Omega_0 + \Omega_1 \cos \left(\Omega_2 t\right)$. Recall that the effective NV-nucleus coupling rate is reduced by a factor of $J_1(\frac{\Omega_1}{\Omega_2}) \approx \frac{\Omega_1}{2 \Omega_2}$ from $g$ to $\approx \frac{\Omega_1}{2 \Omega_2} g$ so the cycle time is increased to $T_{det} = \frac{4 \Omega_2\pi}{\Omega_1 g}$. For an amplitude modulated driving field 
\begin{equation}
P_{AM}^{peak} \propto \left(\Omega_0+\Omega_1\right)^2,\qquad P_{AM}^{cycle} \propto \frac{1}{2}\left(\Omega_0^2+\frac{1}{2}\Omega_1^2\right) T_{AM}.
\end{equation}
Assuming, for example, that $\Omega_2 = 9 \Omega_0$ so $\omega_l = 10 \Omega_0$, which corresponds to the large frequency mismatch, we have that  
\begin{equation}
\frac{P_{HH}^{peak}}{P_{AM}^{peak}} = \frac{\Omega^2}{ \left(\Omega_0+\Omega_1\right)^2}=25,\qquad \frac{P_{HH}^{cycle}}{P_{AM}^{cycle}} \approx \frac{\Omega^2 T_{HH}}{\left(\Omega_0^2+\frac{1}{2}\Omega_1^2\right) T_{AM}}=3.7.
\end{equation}    
   
\section{Quantum sensing}
In this section we show how the $\sigma_-,\sigma_+$ operators can be transformed into a $\sigma_x$ (or $\sigma_y$) operators. 
For the case of the low frequency mismatch this can be achieved by adding an extra drive on the NV in the $y$ or $z$ direction in Eq. \ref{eq7} which rotates at $\Omega_2$ (this amounts to $\Omega_s \cos(\omega_0 t)\cos(\Omega_2 t) \sigma_x$). Specifically, consider the following Hamiltonian, 
\begin{eqnarray}
H &=& \frac{\omega_0}{2} \sigma_z + \frac{\omega_l}{2} I_z + g \sigma_z I_x  \nonumber \\
 &+&  \Omega_1 \sigma_x \cos \left( \omega_0 t \right) - \Omega_2 \sigma_z \cos \left( \Omega_1 t \right) \nonumber \\
 &+& \Omega_s \cos \left( \omega_0 t \right)\cos \left( \Omega_2 t \right) \sigma_x.
 \label{eq_sens1}
\end{eqnarray}
We proceed in a similar manner as in the previous sections. Moving to the IP with respect to $H_0=\frac{\omega_0}{2} \sigma_z $ and to the basis of the dressed states we have that
\begin{eqnarray}
H_I &=& \frac{\Omega_1}{2} \sigma_z + \frac{\omega_l}{2} I_z - g \sigma_x I_x  \nonumber \\
 &+& \Omega_2 \sigma_x \cos \left( \Omega_1 t \right) + \frac{\Omega_s}{2} \cos \left( \Omega_2 t \right) \sigma_z.
\end{eqnarray}
Moving to the second IP with respect to $H_0=\frac{\Omega_1}{2} \sigma_z $ and to the basis of the double-dressed states results in 
\begin{eqnarray}
H_{II} &=& \frac{\Omega_2}{2} \sigma_z + \frac{\omega_l}{2} I_z 
- g \left(\cos\left(\Omega_1 t\right) \sigma_z - \sin\left(\Omega_1 t \right) \sigma_y\right) I_x  \nonumber \\
 &-& \frac{\Omega_s}{2} \cos \left( \Omega_2 t \right) \sigma_x.
\end{eqnarray}
Moving to the third IP with respect to $H_0=\frac{\Omega_2}{2} \sigma_z $ and to the basis of the triple-dressed states results in 
\begin{eqnarray}
H_{III} &\approx & \frac{\Omega_s}{2} \sigma_z + \frac{\omega_l}{2} I_z \nonumber \\
&+& g \sin\left(\Omega_1 t\right) \left(\sin\left(\Omega_2 t\right) \sigma_z + \cos\left(\Omega_2 t \right) \sigma_y\right) I_x,  
\end{eqnarray}
where we have omitted fast rotating terms.  
Thus, in the fourth IP, with respect to $H_0=\frac{\Omega_s}{2} \sigma_z + \frac{\omega_l}{2} I_z$  we get a Hamiltonian which can be used for sensing the nucleus frequency, i.e
\begin{equation}
H_{IV} \approx \frac{g}{4} \sigma_z \left( I_x \cos(\delta t) - I_y \sin(\delta t)   \right),
\label{eq_sens5}
\end{equation}
where $\delta = \Omega_1 +\Omega_2 -\omega_l.$  As the extra term acts as a spin locking term at $\Omega_s,$ the robustness of the methods is not decreased.

The same can be done in the large frequency mismatch regime.
Thus, the interaction should be changed from the flip - flop interaction  $g \left( \sigma_+ I_- + \sigma_- I_+  \right)$ to $g \sigma_x I_x$ by adding, for example,  a $\sigma_x$ drive to the modulation. In this case the Hamiltonian is transformed to (the derivation is similar to the derivation in the small frequency mismatch, Eq. \ref{eq_sens1} -  \ref{eq_sens5})
\begin{equation}
H \approx g J_1\left(\frac{\Omega_1}{\Omega_2} \right) \sigma_x \left(I_x \cos(\delta t) - I_y \sin(\delta t)    \right).
\label{sens1}
\end{equation}

\end{document}